\documentclass[]{fairmeta}

\usepackage{amsmath}
\usepackage{pifont}
\newcommand{\cmark}{\ding{51}}
\newcommand{\xmark}{\ding{55}}

\title{OmniFaceRig: Fully Automatic Inner-Mouth-Aware Face Rigging Across Diverse 3D Character Topologies}

\author[\dagger,\ddagger,*]{Chao Wang}
\author[\dagger]{Guangyao Ma}
\author[\dagger]{John Doublestein}
\author{Junming Chen}
\author{Yiming Lin}
\author{Zhaoen Su}
\author{Xiaomin Luo}
\author{Yushan Yin}
\author{Shiyang Cheng}
\author{Jie Shen}
\author{Doug Roble}
\author{Dilin Wang}
\author[\ddagger]{Yilei Li}
\author[\ddagger]{Rakesh Ranjan}

\affiliation{Reality Labs, Meta}

\contribution[\dagger]{Equal contribution}
\contribution[\ddagger]{Project lead}
\contribution[*]{Corresponding author}

\abstract{
Facial rigging---creating Facial Action Coding System (FACS) blendshapes together with inner-mouth geometry (teeth, gums, and tongue)---remains a major bottleneck in 3D character production. Existing pipelines still depend on substantial designer effort: inner-mouth components are commonly placed by hand, and landmark annotation and per-character template adjustments are routine. We present \textbf{OmniFaceRig}, a fully automatic end-to-end pipeline that converts a static surface-only 3D character mesh, with no pre-modeled oral cavity, into an inner-mouth-aware FACS rig with up to 155 blendshapes, procedurally fitted teeth, gums, and tongue, and re-packed UV/texture. OmniFaceRig supports diverse topologies, including humans, humanoids, long-muzzled animals (elongated snouts; e.g., dogs, wolves, foxes), and short-muzzled animals (compact or nearly absent snouts; e.g., cats, bears, rabbits, tigers), with no manual landmarks, no user-provided templates, and no per-asset template setup. OmniFaceRig combines a hybrid vision-language model (VLM) and computer vision (CV) riggability checker, a four-model face detection and segmentation ensemble (face landmark detection, Sapiens, SAM~3, and fine-tuned Sapiens), dense keypoint-driven template registration, procedural inner-mouth construction, and collision-aware blendshape transfer. To handle non-human characters, it automatically selects from a small library of topology-specific templates with a nose-landmark-free design and uses as-rigid-as-possible (ARAP) deformation with signed-distance-field (SDF) constraints for inner-mouth fitting to reduce teeth--face intersections. We also introduce \textbf{Omni-Bench}, an open-source benchmark of more than 1,000 biped 3D characters rigged with FACS facial blendshapes and inner-mouth geometry, spanning realistic humans, humanoids, cats, dogs, and other animals. Experiments show high final rigging success on screened Omni-Bench candidate inputs, nearly complete face detection recall from the segmentation ensemble, reliable inner-mouth placement with low penetration, and 20--30\,s end-to-end latency on a single A100 GPU including data I/O. Together, OmniFaceRig provides an automatic path from static generated characters to animation-ready facial rigs across both human and non-human topologies. Additional results and resources are available on our project page.
}

\date{June 9, 2026 (v1); October 29, 2026 (v2); November 4, 2026 (v3)}
\metadata[Project Page]{\url{https://omnifacerig.github.io/}}
\metadata[Dataset]{\url{https://github.com/facebookresearch/omnifacerig}}
\metadata[Published in]{ACM Transactions on Graphics 45(6), Article 231 (SIGGRAPH Asia 2026), \href{https://doi.org/10.1145/3842582}{doi:10.1145/3842582}}
\metadata[Keywords]{auto-rigging, facial blendshapes, FACS, benchmark dataset, deformation transfer}

\begin{document}

\maketitle

\begin{figure}[htbp]
  \centering
  \includegraphics[width=0.85\textwidth]{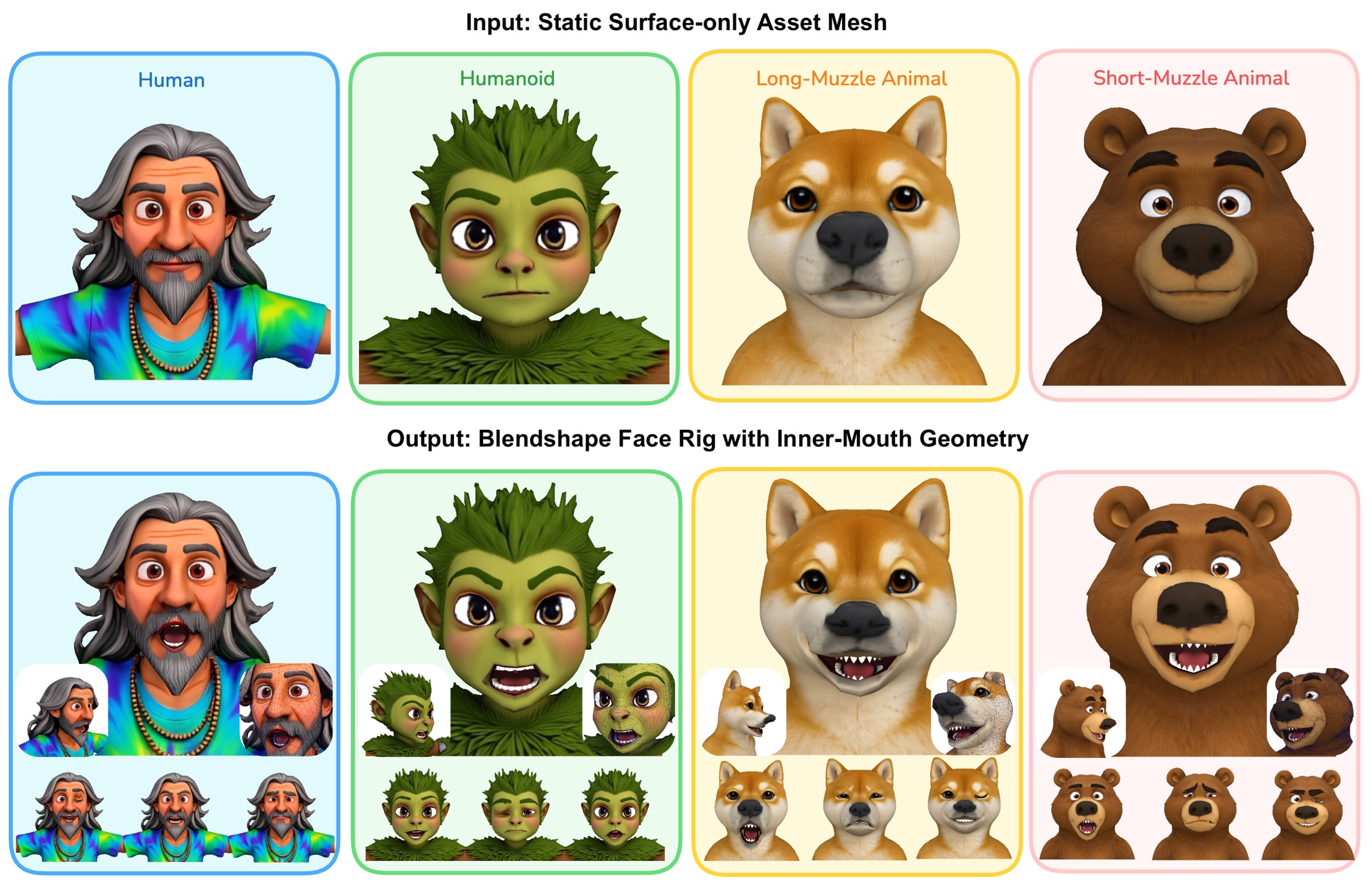}
  \caption{OmniFaceRig converts static surface-only meshes with no oral cavity (top) into production-ready FACS rigs with generated teeth, gums, and tongue (bottom). A single pipeline supports humans, humanoids, and long- and short-muzzled animals.}
  \label{fig:teaser}
\end{figure}

\section{Introduction}
\label{sec:intro}

Facial rigging---creating Facial Action Coding System (FACS) blendshapes and inner-mouth geometry (teeth, gums, and tongue) for 3D character animation---remains one of the most labor-intensive steps in the character production pipeline. A skilled technical artist may spend hours to days sculpting blendshapes for a single character, involving careful topology design, facial landmark placement, expression sculpting, and teeth fitting.

At the same time, 3D asset creation itself has become dramatically more accessible. Recent text-to-3D and image-to-3D methods can generate textured objects or characters directly from prompts or reference images~\citep{poole2023dreamfusion, lin2023magic3d, jun2023shape, liu2023zero123, tang2024make, ren2023make, wang2026assetgen}, and similar capabilities are increasingly available in commercial 3D-generation tools. However, these systems primarily produce static surface meshes and textures. For characters, the generated assets are often plausible as visual meshes but are not animation-ready facial rigs: they typically lack FACS controls, robust facial topology for deformation, and the oral-cavity geometry required for inner-mouth-aware expressions. This makes generated characters natural inputs to an automatic rigging pipeline, but not substitutes for such a pipeline.

Existing rigging solutions address parts of this problem but remain incomplete and, in practice, only \emph{semi-automatic}. Deformation transfer~\citep{sumner2004deformation} and parametric face models~\citep{cao2014facewarehouse, li2017learning} enable facial rig creation but assume compatible mesh topologies, require manual landmark annotation, and are designed exclusively for human faces. Inner-mouth geometry (teeth, gums, and tongue) is particularly costly to author because the input mesh normally contains no corresponding interior surfaces: in production pipelines, teeth, gums, and tongue are almost universally placed and tuned by hand, and manually tuned or human-specific teeth templates routinely fail on stylized or non-human faces. To our knowledge, \emph{no existing system delivers a fully automatic, end-to-end pipeline that produces FACS blendshapes with newly generated inner-mouth geometry from a static surface-only mesh}---let alone across both human and non-human topologies.

We present \textbf{OmniFaceRig}, an end-to-end pipeline for face auto-rigging \textbf{across diverse 3D character topologies}---including humans, humanoids, long-muzzled animals (e.g., dogs, wolves, foxes), and short-muzzled animals (e.g., cats, bears, rabbits, tigers). These animal labels describe coarse facial topology for template selection rather than biological taxonomy: long-muzzled characters have elongated, forward-protruding snouts, while short-muzzled characters have compact or nearly absent snouts. Given a static surface-only 3D mesh with no pre-modeled oral cavity, OmniFaceRig produces a production-ready facial rig with up to 155 FACS blendshapes, procedurally generated inner-mouth geometry (teeth, gums, and tongue), and re-packed UV/texture---all \textbf{fully automatic with no manual intervention at any stage}. In particular, OmniFaceRig eliminates the manual inner-mouth placement, landmark annotation, user-provided templates, and per-asset template setup that bottleneck prior pipelines, allowing the entire facial rig to be generated directly from the input mesh (\Cref{fig:teaser}). Our key contributions are:

\begin{enumerate}
  \item To our knowledge, the first \textbf{fully automatic, end-to-end face auto-rigging pipeline} that starts from a static surface-only input mesh and requires no manual inner-mouth placement, landmark annotation, user-provided templates, or per-asset template setup.
  \item \textbf{Inner-mouth-aware rig construction}: OmniFaceRig procedurally generates teeth, gums, tongue, and oral-cavity support geometry that are absent from the input mesh, then integrates them with FACS blendshape transfer using as-rigid-as-possible (ARAP) deformation, signed-distance-field (SDF) constraints, and collision-aware deformation.
  \item \textbf{Omni-Bench}: to our knowledge, the first open-source benchmark of more than 1,000 biped 3D characters with FACS facial blendshapes and inner-mouth geometry (teeth, gums, and tongue), spanning realistic humans, humanoids, domestic cats and dogs, and wild animals---all in T-pose and all rigged by OmniFaceRig.
  \item Generalization across \textbf{diverse non-human topologies}: our method rigs long-muzzled animals (dogs, wolves, foxes, etc.) and short-muzzled animals (cats, bears, rabbits, tigers, etc.) via a small library of topology-specific templates with a nose-landmark-free design and ARAP+SDF teeth placement, demonstrating that the same pipeline transfers from humans to animals without architectural changes.
\end{enumerate}

\section{Related Work}
\label{sec:related}

\begin{table}[tbp]
\centering
\caption{Comparison of facial auto-rigging methods. OmniFaceRig is the only compared method achieving fully automatic blendshape generation with no per-asset setup, supporting disconnected components including inner-mouth geometry (teeth, gums, and tongue), and generalizing across diverse human and animal character topologies. ``No Per-Asset Setup'' means no user-provided template, manual landmarking, or per-character fitting adjustment at inference time.}
\label{tab:method_comparison}
\resizebox{\textwidth}{!}{
\begin{NiceTabular}{lccccccc}
\CodeBefore
  \rowcolor{metabg}{10}
\Body
\toprule
\textbf{Method} & \Block{}{\textbf{Cross-}\\\textbf{Topology}} & \Block{}{\textbf{No Manual}\\\textbf{Landmarks}} & \Block{}{\textbf{No Per-Asset}\\\textbf{Setup}} & \Block{}{\textbf{Output}\\\textbf{Format}} & \Block{}{\textbf{Disconnected}\\\textbf{Parts}} & \Block{}{\textbf{Inner}\\\textbf{Mouth}} & \Block{}{\textbf{Animal}\\\textbf{Support}} \\
\midrule
Deformation Transfer~\citep{sumner2004deformation} & \xmark & \xmark & \xmark & Mesh & \xmark & \xmark & \xmark \\
Example-Based~\citep{li2010example} & \xmark & \xmark & \xmark & Blendshapes & \xmark & \xmark & \xmark \\
\citet{li2020dynamic} & \xmark & \xmark & \xmark & Blendshapes & \xmark & \xmark & \xmark \\
\citet{chandran2022shape} & \cmark & \xmark & \xmark & Mesh & \xmark & \xmark & \xmark \\
NFR~\citep{qin2023neural} & \cmark & \cmark & \cmark & Mesh / Latent & \xmark & \xmark & \xmark \\
NFDT~\citep{chandran2025neural} & \cmark & \cmark & \cmark & Blendshapes & \xmark & \xmark & Limited \\
NFS~\citep{cha2025neural} & \cmark & \cmark & \cmark & Skinning Weights & \cmark & \xmark & \xmark \\
RigAnyFace~\citep{ma2025riganyface} & \cmark & \cmark & \cmark & Blendshapes & \cmark & \xmark & \xmark \\
\midrule
\textbf{OmniFaceRig (Ours)} & \cmark & \cmark & \cmark & \textbf{Blendshapes} & \cmark & \cmark & \cmark \\
\bottomrule
\end{NiceTabular}
}
\end{table}

\begin{figure}[tbp]
  \centering
  \includegraphics[width=\textwidth]{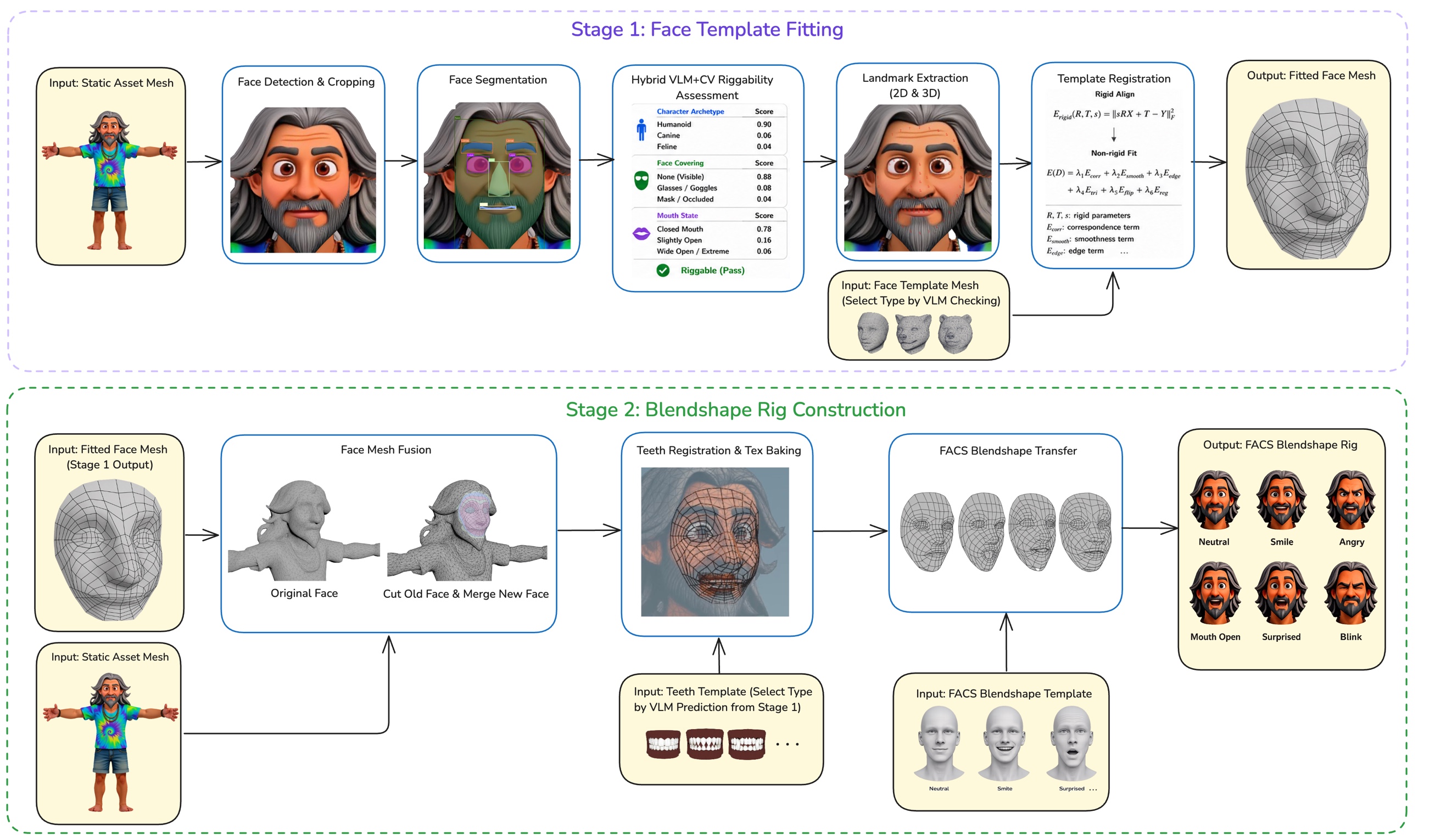}
  \caption{OmniFaceRig pipeline architecture. The pipeline consists of two stages. \textbf{Stage~1: Face Template Fitting} takes a static surface-only 3D mesh, performs face detection/cropping, face segmentation, hybrid vision-language model (VLM) and computer vision (CV) riggability assessment, landmark extraction, and anchor-driven rigid + non-rigid template registration, producing a fitted template mesh aligned to the input face. \textbf{Stage~2: Blendshape Rig Construction} takes the fitted template and constructs a production-ready FACS blendshape rig through face mesh fusion, generated teeth registration and texture baking, and FACS blendshape transfer, producing inner-mouth geometry (teeth, gums, and tongue) and re-packed UV/texture that are not present in the input.}
  \label{fig:pipeline}
\end{figure}

\subsection{3D Asset and Character Generation}

Text-to-3D and image-to-3D generation have rapidly lowered the barrier to creating 3D assets. Methods such as DreamFusion~\citep{poole2023dreamfusion}, Magic3D~\citep{lin2023magic3d}, Shap-E~\citep{jun2023shape}, and Zero-1-to-3~\citep{liu2023zero123} generate textured objects from language or image inputs, while character-focused systems such as Make-It-Vivid~\citep{tang2024make} and Make-A-Character~\citep{ren2023make} target animatable or stylized 3D characters. Similar capabilities are also increasingly available in commercial 3D-generation services, making static character asset creation accessible to non-expert users. However, these systems primarily target visible geometry and appearance. Their outputs are generally static surface assets, often without production facial topology, FACS controls, or interior oral-cavity geometry. Our work is complementary: OmniFaceRig takes such generated surface meshes as input and converts them into animation-ready, inner-mouth-aware facial rigs.

\subsection{Parametric Face Models vs.\ Blendshape Rigging}

Parametric face models provide statistical representations of human facial geometry. The 3D Morphable Model (3DMM)~\citep{blanz1999morphable}, FaceWarehouse~\citep{cao2014facewarehouse}, and FLAME~\citep{li2017learning} established linear shape spaces for face reconstruction. Extensions include multilinear models~\citep{brunton2014multilinear, vlasic2006face} that factor identity and expression, and disentangled representations~\citep{abrevaya2019decoupled, jiang2019disentangled} that separate shape components for more flexible control. Methods like CoMA~\citep{ranjan2018generating} and DECA~\citep{feng2021learning} leverage these models to reconstruct animatable faces from images. However, parametric models are bound to a fixed topology and blendshape set, and cannot be applied to novel characters with arbitrary topologies, stylized proportions, or non-human features.

In contrast, production pipelines rely on FACS~\citep{ekman1978facial} and blendshape rigging~\citep{lewis2014practice, lewis2010direct}, which offer animators intuitive, localized control while preserving character-specific geometry~\citep{seol2011artist, kavan2024compressed}. \citet{orvalho2012facial} provide a comprehensive survey of facial rigging techniques, covering the evolution from manual sculpting to semi-automatic methods. On the representation side, spectral methods have also been explored for compact animation representations~\citep{wang2015spectral}. Our method follows this production paradigm, generating FACS blendshapes for arbitrary target meshes.

In conventional digital-content-creation workflows using tools such as Maya or MotionBuilder, technical artists typically prepare animation-friendly facial topology, align landmarks or templates, place and adjust teeth, gums, and tongue, sculpt or transfer expression shapes, and manually correct collisions. OmniFaceRig targets these repeated geometric preparation and cleanup steps while retaining the standard FACS blendshape representation used by downstream animation tools.

\subsection{Traditional and Learning-Based Facial Auto-Rigging}

\textbf{Traditional methods.} \citet{noh2001expression} introduced expression cloning via vertex displacement transfer. \citet{sumner2004deformation} proposed deformation transfer using triangle affine transformations---the foundational technique our system builds upon. \citet{orvalho2008transferring} extended this to transfer rigs across different face models. \citet{li2010example} further developed example-based facial rigging. \citet{seol2012spacetime} introduced spacetime expression cloning for blendshapes. \citet{liu2011framework} proposed a framework for locally retargeting and rendering facial performance. \citet{ribera2017facial} addressed retargeting with automatic range of motion alignment. These methods achieve high quality but require manual correspondence establishment, which does not scale.

\textbf{Learning-based methods.} \citet{li2020dynamic} used CNNs to predict blendshape offsets via geometry images. \citet{chandran2020semantic} introduced semantic deep face models, later used transformers for topology-independent shape modeling \citep{chandran2022shape}, and proposed anatomically-constrained facial performance retargeting \citep{chandran2022local}. Blendshape editing has also been explored through sketching interfaces \citep{cetinaslan2020sketching} and localized Jacobian transpose descent \citep{cetinaslan2020stabilized}. \citet{wang2023fully} proposed a method for fully automatic blendshape generation for stylized characters, and \citet{ming2024high} introduced high-quality mesh blendshape generation from face videos via neural inverse rendering. On the representation side, MeshCNN \citep{hanocka2019meshcnn} and DiffusionNet \citep{sharp2022diffusionnet} provide mesh-native learned features that benefit downstream tasks. Neural Face Rigging (NFR) \citep{qin2023neural} pioneered neural deformation for in-the-wild facial meshes but is limited to human faces and cannot handle disconnected components. Fabrig \citep{zhu2024fabrig} proposes a cloth-simulated transferable 3D face parameterization. Neural Facial Deformation Transfer (NFDT) \citep{chandran2025neural} uses a decoder-only transformer for high-fidelity transfer. Neural Face Skinning (NFS) \citep{cha2025neural} proposes mesh-agnostic skinning. RigAnyFace \citep{ma2025riganyface} represents the current state of the art, using 2D supervision to scale training with unlabeled data. Most recently, CANRig \citep{canrig2026} introduces cross-attention mechanisms for neural face rigging with variable local control, and ControlFace \citep{jang2025controlface} harnesses 3DMM-conditioned rendering for flexible face rigging. Zero-shot approaches \citep{wang2023zeroshot} and image-space portrait and talking-face animation methods \citep{yang2024megactor,xie2024xportrait,zhang2025magictalk} have also been explored for stylized characters, and dyadic audiovisual models generate speech-aligned gestures and expressions~\citep{agrawal2025seamless}. DreamFace \citep{zhang2023dreamface} generates animatable 3D faces under text guidance. As summarized in \Cref{tab:method_comparison}, our method addresses critical gaps: inner-mouth modeling (teeth, gums, and tongue) and non-humanoid character rigging.

\subsection{Rigging for Non-Humanoid Characters}

Prior work has explored automatic generation and stylization of facial rigs~\citep{danieau2019automatic}. Extending facial rigging to animals requires handling substantially different anatomy. On the 3D shape modeling side, SMAL~\citep{zuffi2017menagerie} and BARC~\citep{rueegg2023barc} model quadruped animal shape and pose using category-specific priors. However, these parametric animal models do not address facial rig construction or blendshape generation. For facial landmark detection on non-photorealistic characters, \citet{cheng2024stylizedfacepoint} proposed methods for stylized character facial landmark detection, addressing the domain gap between real human faces and artistic character designs. Recent learning-based rigging methods~\citep{qin2023neural, ma2025riganyface} are trained on human datasets with human-centric priors and do not generalize to animal faces. Critically, nearly all existing methods focus on the outer facial surface, ignoring the oral cavity (teeth, gums, tongue). Our approach introduces a robust multi-model segmentation ensemble that identifies facial regions across species, decoupling geometric processing from human-specific priors. OmniFaceRig generates FACS blendshapes that explicitly include inner-mouth geometry (teeth, gums, and tongue) for both humans and animals.

\section{Omni-Bench: Benchmark Dataset}
\label{sec:benchmark}

To facilitate future research in automatic facial rigging, we release \textbf{Omni-Bench}, to our knowledge the first open-source benchmark of biped 3D characters with FACS facial blendshapes and inner-mouth geometry (teeth, gums, and tongue). The source characters are static surface meshes generated with AssetGen~\citep{wang2026assetgen} from T-pose images, each produced from a text prompt. All released assets are selected from inputs that pass the initial VLM+CV riggability screen, are processed by OmniFaceRig, and are provided in biped T-pose with generated rig outputs.

\begin{table}[tbp]
\centering
\caption{Comparison of 3D facial and character datasets. To our knowledge, Omni-Bench is the first dataset providing FACS blendshapes with inner-mouth geometry (teeth, gums, and tongue) for both human and animal full-character assets, along with the text prompt and 2D input image of every asset.}
\label{tab:dataset_comparison}
\resizebox{\textwidth}{!}{
\begin{NiceTabular}{lccccccc}
\CodeBefore
  \rowcolor{metabg}{10}
\Body
\toprule
\textbf{Dataset} & \textbf{Year} & \Block{}{\textbf{Total}\\\textbf{Models}} & \textbf{Species} & \Block{}{\textbf{FACS}\\\textbf{Blendshapes}} & \Block{}{\textbf{Inner}\\\textbf{Mouth}} & \Block{}{\textbf{Full}\\\textbf{Character}} & \Block{}{\textbf{Text +}\\\textbf{2D Image}} \\
\midrule
BFM~\citep{paysan20093d} & 2009 & 200 & Human & \xmark & \xmark & \xmark & \xmark \\
CoMA~\citep{ranjan2018generating} & 2018 & 144 & Human & \xmark & \xmark & \xmark & \xmark \\
VOCASET~\citep{cudeiro2019capture} & 2019 & 12 (4D) & Human & \xmark & \xmark & \xmark & \xmark \\
FaceScape~\citep{yang2020facescape} & 2020 & 16,940 & Human & \xmark & \xmark & \xmark & \xmark \\
ICT FaceKit~\citep{li2020learning} & 2020 & 1 (Parametric) & Human & \cmark & \cmark & \xmark & \xmark \\
Multiface~\citep{wuu2022multiface} & 2022 & 13 (Multi-view) & Human & \xmark & \xmark & \xmark & \xmark \\
RaBit~\citep{luo2023rabit} & 2023 & 1,500 & Human \& Cartoon & \xmark & \xmark & \cmark & \xmark \\
Anymate~\citep{anymate2025} & 2025 & 230,000 & Human \& Animal & \xmark & \xmark & \cmark & \xmark \\
\midrule
\textbf{Omni-Bench (Ours)} & 2026 & \textbf{1,000+} & \textbf{Human \& Animal} & \cmark & \cmark & \cmark & \cmark \\
\bottomrule
\end{NiceTabular}
}
\end{table}

Existing facial datasets leave complementary gaps. BFM~\citep{paysan20093d} is a parametric model, while FaceScape~\citep{zhu2023facescape, yang2020facescape}, Multiface~\citep{wuu2022multiface}, and VOCASET~\citep{cudeiro2019capture} focus on human outer-face geometry or captured motion rather than complete FACS rigs. ICT FaceKit~\citep{li2020learning} includes oral components but is a single parametric model. Large character collections such as RaBit~\citep{luo2023rabit} and Anymate~\citep{anymate2025} lack both FACS facial blendshapes and inner-mouth structures.

\begin{figure}[p]
  \centering
  \includegraphics[width=\textwidth,height=0.88\textheight,keepaspectratio]{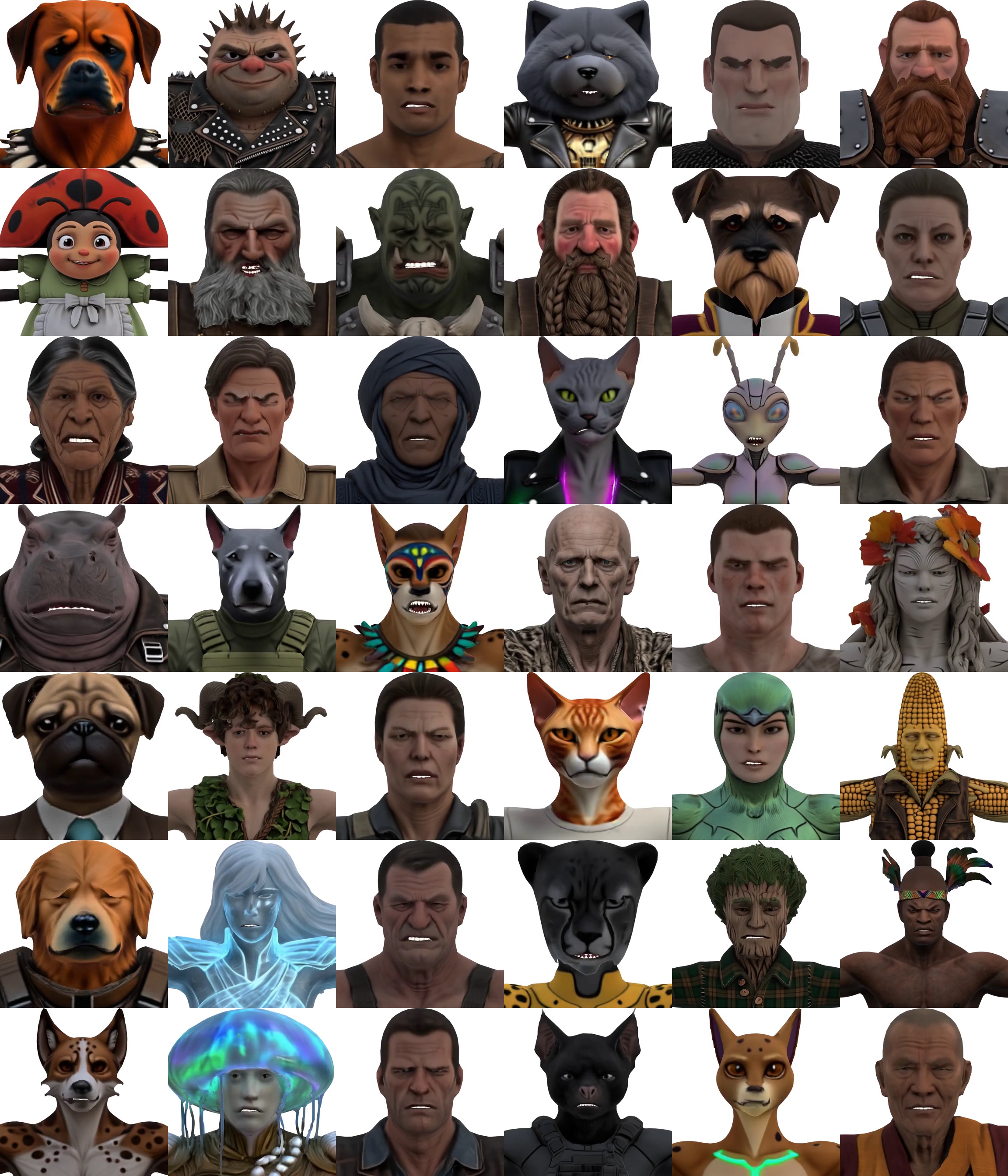}
  \caption{\textbf{Omni-Bench overview.} Representative rigged characters with randomly sampled FACS expressions, spanning humans, humanoids, cats, dogs, and other animals.}
  \label{fig:omnibench_overview}
\end{figure}

Omni-Bench contains \textbf{more than 1,000 rigged biped 3D characters}; the public v1 release provides 1,345 assets, comprising 809 realistic or stylized humans/humanoids and 536 animals. They span human professions, fantasy races, science fiction, nature and elemental humanoids, domestic cats and dogs, and wild felines and canines (281 short- and 255 long-muzzled faces). \Cref{fig:omnibench_overview} summarizes this diversity. The rig-quality experiments use smaller subsets with independent artist-authored references (\Cref{subsec:metrics}).

The riggability screen should be interpreted as an input eligibility filter rather than a guarantee of final rigging success. It removes obvious invalid assets---for example, missing faces, severe eye or mouth occlusions, non-standard viewpoints, or unsupported facial topologies---before full processing. A screened asset can still fail during later stages because of segmentation errors, unstable template fitting, or implausible mouth localization. We therefore report final \emph{Success Rate} separately in the experiments (\Cref{subsec:metrics}) as the percentage of screened Omni-Bench candidate inputs for which the full OmniFaceRig pipeline completes and passes final geometric quality checks. Success Rate is measured before unsuccessful outputs are excluded, rather than on the released assets alone.

As shown in \Cref{tab:dataset_comparison}, Omni-Bench uniquely combines human and animal full-character assets with up to 155 auto-generated FACS shapes with teeth, gums, and tongue, and the complete generation provenance (text prompt, 2D input image, and final mesh). All assets are in biped T-pose, processed by the same pipeline, and also released with segmentation and keypoint images, solved templates, and FBX/glTF/USDZ rigs.

\section{Method}
\label{sec:method}

\subsection{Pipeline Overview}
\label{sec:overview}

OmniFaceRig takes a static surface-only 3D mesh with no pre-modeled oral cavity and produces a production-ready FACS rig with teeth, gums, and tongue (\Cref{fig:pipeline}). \textbf{Stage~1: Face Template Fitting} (\Crefrange{sec:riggability}{sec:fitting}) combines face localization, multi-model segmentation, hybrid VLM+CV riggability assessment, keypoint extraction, and rigid/non-rigid registration to align a topology-specific template to the input. \textbf{Stage~2: Blendshape Rig Construction} (\Cref{sec:blendshape}) fuses the fitted face with the asset, generates and registers inner-mouth geometry, rebuilds its UV texture, and transfers the canonical FACS blendshapes.

\subsection{Riggability Assessment}
\label{sec:riggability}

Before processing an asset through the rigging pipeline, OmniFaceRig performs an automatic eligibility check. This is critical in production settings where input assets are generated automatically (e.g., from text-to-3D models) and may include characters with no visible face, heavily occluded facial features, non-standard orientations, or facial topologies outside the current template library.

Our riggability assessment is a \textbf{VLM+CV hybrid} system combining two complementary approaches:

\textbf{VLM Understanding.} We use Gemini 2.5 Flash Image as the primary vision-language model (VLM), with a Llama 4 multimodal model as a fallback when the primary response is unavailable or invalid. We render the input 3D mesh from a frontal viewpoint and query the VLM to identify: (1) the \emph{character archetype} (human, humanoid, feline, canine, or other); (2) \emph{face covering status} (glasses, goggles, masks, eye patches); and (3) \emph{teeth state} (open mouth, closed mouth, emotional expression). The VLM's flexibility allows easy adaptation to new assessment criteria without retraining. The VLM has two roles. First, it filters out assets that are unlikely to be riggable by the current system, including no-face assets, faces with severe eye or mouth occlusion, and unsupported facial topologies such as highly exaggerated beaked birds or other shapes outside our human, long-muzzle, and short-muzzle template families. Second, for assets that pass this eligibility check, it selects downstream configuration choices: (i) the \emph{face template variant}---one of \emph{human}, \emph{long-muzzle}, or \emph{short-muzzle}---based on the character's facial topology (see \Cref{sec:fitting}); and (ii) the \emph{teeth archetype} (human, canine, or monster), which Stage~2 (\Cref{sec:blendshape}) uses for inner-mouth fitting.

\textbf{CV Segmentation Filtering.} After face localization, we compute face segmentation masks using the multi-model ensemble described in \Cref{sec:segmentation}. Assets where no valid face region is detected, or where the face area falls below a minimum threshold, are filtered out. This CV-based check catches cases where the VLM may be overconfident---for example, detecting a ``face'' on a purely mechanical character with no actual facial geometry.

\textbf{Hybrid Decision.} The final eligibility decision combines VLM predictions with CV filtering results. We evaluate the hybrid approach on a mixed riggability validation set of approximately 500 assets, including Omni-Bench candidates, a subset of a separate benchmark collection, and additional generated or collected examples. This validation set intentionally includes both riggable and non-riggable inputs; metrics compare the automatic checker decision against human ground-truth riggability labels (\Cref{tab:riggability}).

\begin{table}[tbp]
  \centering
  \caption{Riggability assessment accuracy on a mixed validation set of approximately 500 assets with human ground-truth riggability labels. The hybrid VLM+CV approach achieves the best performance.}
  \label{tab:riggability}
  \begin{NiceTabular}{lccc}
\CodeBefore
  \rowcolor{metabg}{4}
\Body
    \toprule
    Method & Accuracy & Recall & F1 \\
    \midrule
    VLM only & 94.69\% & 94.76\% & 94.71\% \\
    CV segmentation filtering only & 76.72\% & 94.87\% & 80.87\% \\
    \textbf{VLM + CV (Ours)} & \textbf{$>$95\%} & \textbf{$>$95\%} & \textbf{$>$95\%} \\
    \bottomrule
  \end{NiceTabular}
\end{table}

The hybrid approach leverages the VLM's strong image understanding for high-level decisions while using CV signals for geometric validation, consistently outperforming either component alone. Additional details on the per-question signal assignment, segmentation-mask checks, and mouth-signal fusion are provided in Appendix~\ref{supp:sec:riggability}.

\subsection{Sapiens-Based Segmentation Models}
\label{sec:seg_models}

\subsubsection{Pretraining}

Our learned parsers use Sapiens-1B~\citep{khirodkar2025sapiens}, a high-resolution ViT encoder trained by masked reconstruction on a large human-centric corpus. Its patch-level representation preserves the local boundary cues needed for hairlines, eyelids, lips, teeth, and inner-mouth regions. We adapt this shared encoder to stylized characters before supervised parser training; the full pretraining and adaptation schedule is reported in \Cref{subsec:implementation}.

\subsubsection{Segmentation Fine-Tuning}

We attach a lightweight three-stage deconvolutional decoder and fine-tune it while freezing the ViT backbone. Two checkpoints share this architecture but use different label spaces: a 41-class stylized parser for human-like characters and a 38-class humanoid parser for animals and strongly non-human characters. The former retains fine facial and accessory labels, whereas the latter trades some granularity for stability on cats, dogs, and heavily stylized faces. At inference, their labels are collapsed into the geometric cues required downstream: face boundary, eyes, nose or muzzle extent, mouth opening, inner-mouth support, and occlusion regions. Training details are given in \Cref{subsec:implementation}.

\subsection{Face Detection and Segmentation}
\label{sec:segmentation}

Robust face detection and segmentation are critical prerequisites for template fitting. \textbf{Render and Face Detection.} For each input asset, we render the 3D mesh from a frontal viewpoint and use an in-house face detector and sparse landmark estimator, with MediaPipe~\citep{lugaresi2019mediapipe} as a fallback, to obtain a face bounding box and sparse 2D landmarks. The box is expanded using predefined asymmetric margins to include additional forehead, jaw, and lateral-face context; the enlarged region is then cropped and resized so that the face is approximately centered. Downstream stages consume only the resulting bounding box and landmark coordinates and are therefore not tied to a proprietary landmark representation. The cropped face image is the input to all subsequent face-parsing modules. \textbf{Multi-Model Segmentation.} While face parsing methods such as BiSeNet~\citep{yu2018bisenet} and the RoI Tanh-polar Transformer~\citep{lin2021roi} have achieved strong results on photographic human faces and support downstream tasks such as age estimation~\citep{lin2025fpage}, AI-generated characters exhibit highly stylized proportions and diverse topologies that challenge any single detector or parser. To maximize coverage, we combine the Sapiens-based parsers from \Cref{sec:seg_models} with two complementary models---the in-house face landmark detector described above and the off-the-shelf SAM~3---and adaptively select the best result for each facial region:

\begin{enumerate}
  \item \textbf{Face Landmark Detection}: The in-house detector and sparse landmark estimator described above, with MediaPipe as a fallback. It provides high-precision eye, mouth, and face-boundary cues on human and near-human faces.
  \item \textbf{Sapiens}~\citep{khirodkar2025sapiens}: The base human parser described in \Cref{sec:seg_models}. It provides strong human nose, lip, and facial-skin estimates, but degrades on strongly non-human faces because of its human-centric training distribution.
  \item \textbf{SAM 3} (Segment Anything Model 3)~\citep{carion2025sam3}: Building on the Segment Anything foundation~\citep{kirillov2023segment} and its video extension SAM~2~\citep{ravi2024sam2}, SAM~3 is a class-agnostic segmentation foundation model that provides zero-shot segmentation masks from point or box prompts. SAM 3 generalizes well to stylized and non-human faces but produces coarser boundaries compared to landmark-based methods. We use SAM 3 for face region segmentation on characters where face landmark detection and Sapiens fail.
  \item \textbf{Fine-Tuned Sapiens}: Our stylized and humanoid Sapiens checkpoints from \Cref{sec:seg_models}. The riggability stage selects the 41-class stylized variant for human-like characters and the 38-class humanoid variant for cats, dogs, and heavily stylized assets; these models serve as the primary dense parsers whenever face landmark detection or the base Sapiens model becomes unreliable.
\end{enumerate}

All four models run in parallel, and the system \emph{selects the best result independently for each facial region}:
\begin{itemize}
  \item Humans primarily use landmark cues for eyes/mouth and Sapiens for nose/skin.
  \item Stylized humanoids primarily use the 41-class fine-tuned parser.
  \item Animals primarily use the 38-class parser, with SAM~3 as fallback.
  \item If a preferred model fails, only that region switches to the next-best output.
\end{itemize}

\begin{figure}[tbp]
  \centering
  \begin{subfigure}[b]{0.52\linewidth}
    \centering
    \includegraphics[width=\linewidth]{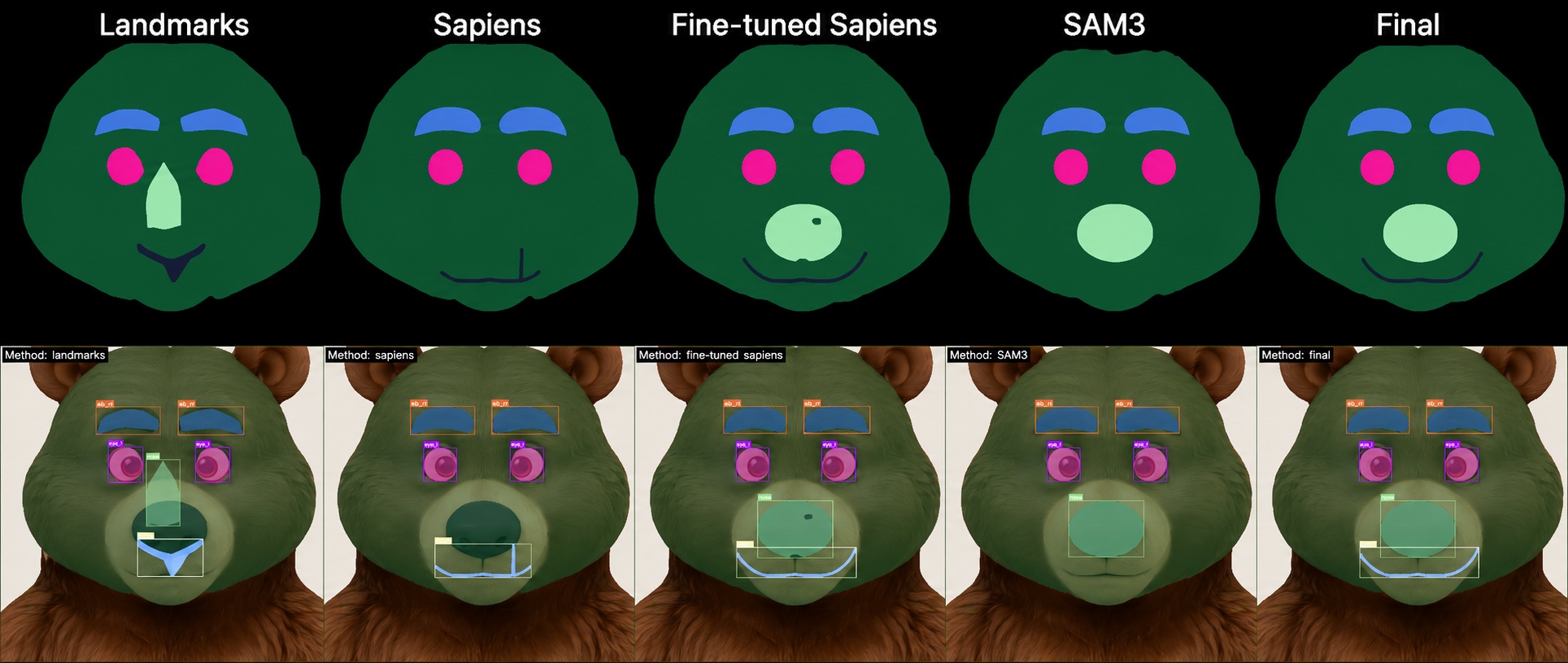}
    \caption{Single-model-dominant case.}
    \label{fig:segmentation_a}
  \end{subfigure}
  \hfill
  \begin{subfigure}[b]{0.44\linewidth}
    \centering
    \includegraphics[width=\linewidth]{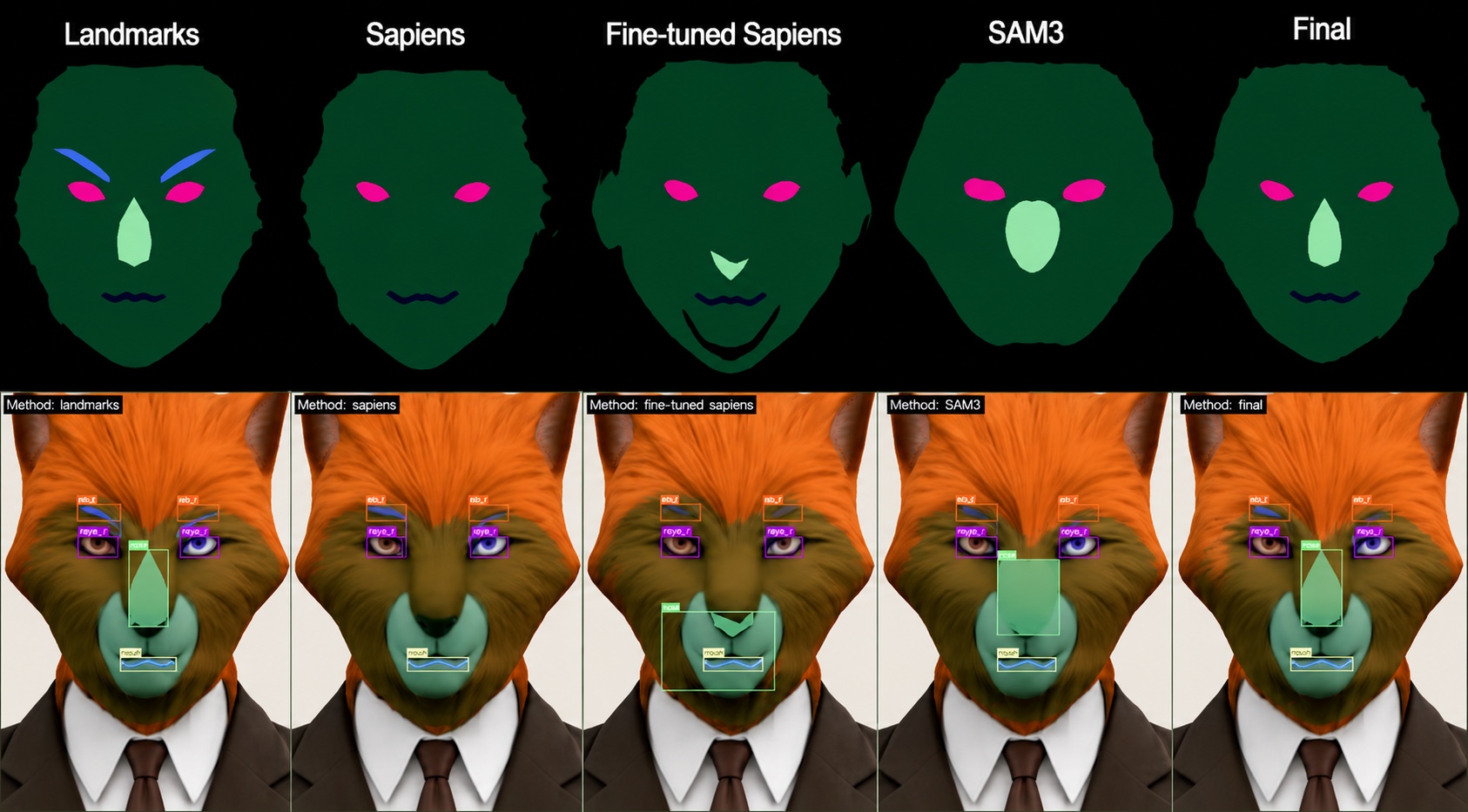}
    \caption{Ensemble case (per-region fusion needed).}
    \label{fig:segmentation_b}
  \end{subfigure}
  \caption{Multi-model face segmentation. \textbf{(a)} Fine-tuned Sapiens dominates when its parse is reliable. \textbf{(b)} When no single model succeeds, the final mask selects the best output independently for each facial region.}
  \label{fig:segmentation}
\end{figure}

The combined multi-model ensemble achieves a face detection recall of \textbf{nearly 100\%} across our test sets, compared to \textbf{88\%} for the best single model in isolation. This improvement is especially pronounced on challenging cases: animal characters with fur-covered face boundaries, humanoid characters with exaggerated feature proportions, and characters wearing partial face accessories. \Cref{fig:segmentation} illustrates single-model and per-region ensemble results.

\textbf{Landmark Extraction (2D \& 3D).} After segmentation, we extract 2D and 3D keypoints from the final segmentation masks. The nose or muzzle mask remains available as a semantic and geometric cue, but it is excluded from the fitting-anchor set described in \Cref{sec:fitting}. The registration correspondences are instead sampled from the face boundary, eyes, and mouth using mask contours and face-landmark predictions. We then back-project each 2D anchor onto the input 3D mesh by ray-casting from the rendered camera through the pixel and intersecting the asset surface, yielding the paired 3D position. The resulting 2D + 3D anchor correspondences are the input to template registration (\Cref{sec:fitting}).

\subsection{Template Registration}
\label{sec:fitting}

\subsubsection{Template Design}

The core of OmniFaceRig's face fitting stage is a small library of quad-mesh templates with anchor keypoints, designed for robust fitting across diverse character topologies. The anchor keypoints correspond to facial landmarks (eyes, mouth, face boundary) whose 3D positions are computed from the segmentation ensemble output (\Cref{sec:segmentation}). To accommodate the wide variation in facial shape across topologies, we provide \textbf{three face template variants}, and the VLM (\Cref{sec:riggability}) automatically selects the appropriate one for each input asset:

\begin{itemize}
  \item \textbf{Human template} --- targets human and humanoid characters with conventional facial proportions (flat-fronted face, defined chin, eyes and mouth on roughly the same plane).
  \item \textbf{Long-muzzle template} --- targets long-muzzled animals with elongated, forward-protruding snouts, such as dogs, wolves, foxes, and horses.
  \item \textbf{Short-muzzle template} --- targets short-muzzled animals with compact or nearly absent snouts, such as cats, bears, rabbits, and tigers.
\end{itemize}

\begin{figure}[tbp]
  \centering
  \includegraphics[width=0.6\linewidth]{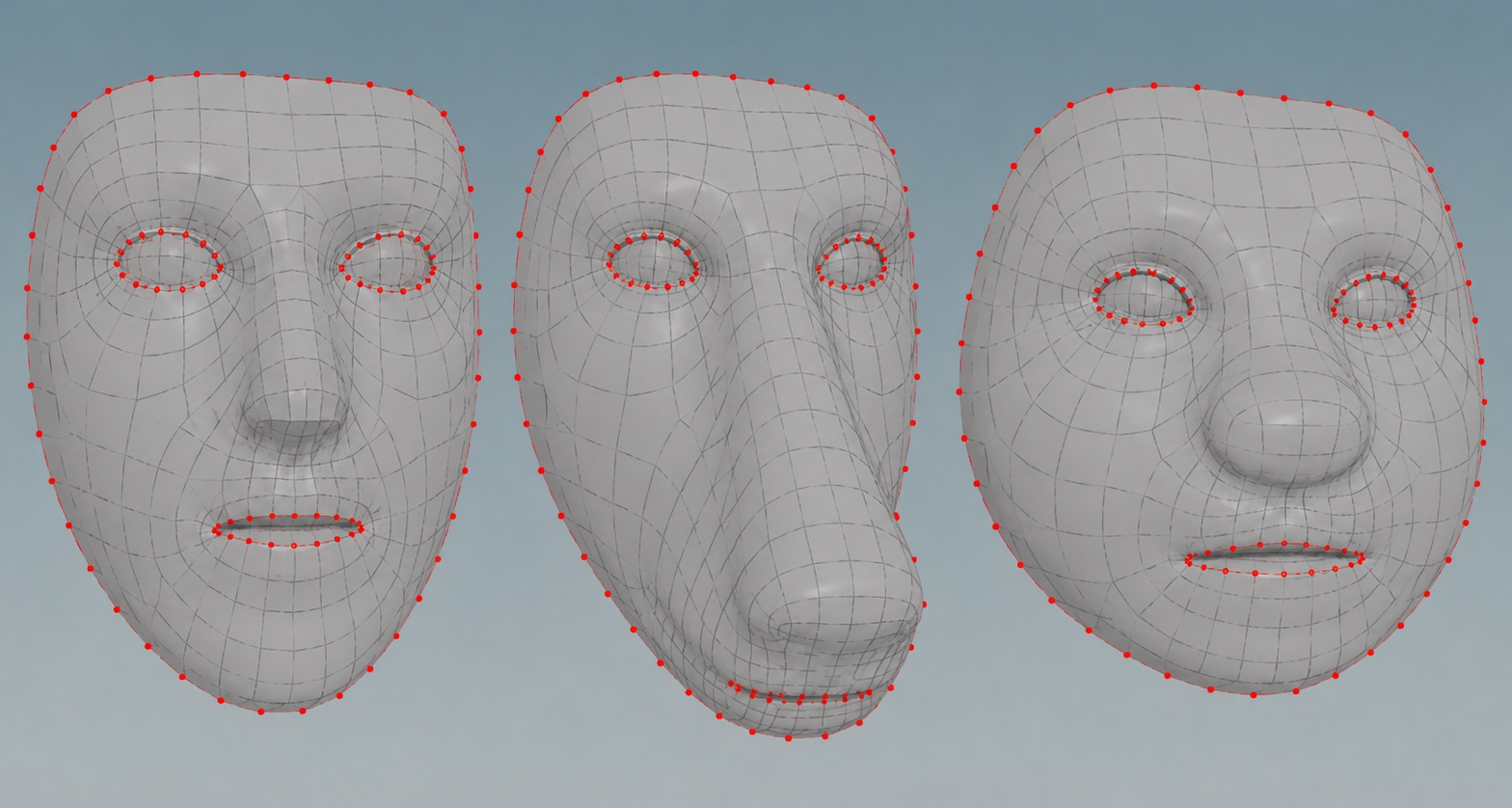}
  \caption{\textbf{Topology-specific face templates.} Human, long-muzzle, and short-muzzle templates (left to right), with fitting anchors marked in red. All templates share the same topology and connectivity.}
  \label{fig:face_templates}
\end{figure}

The three variants differ only in their canonical vertex positions: the human template provides a conventional facial initialization, while the two animal templates initialize the optimization closer to elongated or compact muzzle proportions. Their shared connectivity keeps the downstream fitting and animation pipeline unchanged across categories (\Cref{fig:face_templates}).

All three templates share three common design principles that make fitting robust across diverse character types:

\begin{itemize}
  \item \textbf{Nose-landmark-free design.} Anchor target positions are computed by uniformly sampling along the corresponding segmentation-mask boundary, and nose anchors are particularly fragile under this scheme for two compounding reasons. First, nose segmentation accuracy itself is consistently low on animal faces (feline noses are small and short; canine noses merge into the muzzle) and on many stylized humanoids, so the underlying mask is unreliable. Second, even when the nose mask is plausible, the nose contour tends to be small, irregular, or weakly defined, so uniform boundary sampling yields unstable keypoints that do not correspond across assets. We therefore exclude the nose region from the anchor keypoints entirely; the nose vertices are positioned by the optimization itself, guided by the surrounding border, eye, and mouth anchors. This makes fitting independent of nose segmentation quality, and the resulting nose region naturally adapts to actual face proportions---compact for cats, elongated for dogs, standard for humans.
  \item \textbf{Decoupled fitting and animated topology.} The fitting template (optimized for keypoint correspondence) and the animated topology (designed with radial edge loops around eyes and mouth for muscle-direction deformation) are maintained as separate meshes linked by UV correspondence. This decoupling allows independent optimization: the fitting template can be modified for better landmark matching without affecting the animation-quality edge flow, and vice versa.
  \item \textbf{Minimum region fitting.} The template adheres to the convex hull of facial features rather than extending to the neck and chin. This ``minimum region'' design produces tighter fits, reduces artifacts at the face-asset merge boundary, and improves robustness on characters with unusual neck or chin geometry (e.g., characters with armor, scarves, or elongated chins).
\end{itemize}

\subsubsection{Keypoint Computation and Fallback Mechanisms}

Template fitting is driven by 68 facial anchor correspondences distributed around the face boundary, eyes, and mouth. These sparse anchors constrain the correspondence objective, while all remaining template vertices are optimized through the regularized non-rigid deformation. The target positions are computed from the segmentation ensemble: eye corners and contours from face landmark detection or fine-tuned Sapiens, mouth corners from landmark detection, and face boundary from segmentation mask edges.

To avoid invalid keypoint errors in the evaluated assets, we implement three fallback mechanisms:
\begin{enumerate}
  \item \textbf{Mesh-surface fallback for border keypoints.} After correcting invalid border keypoints to the nearest face mask boundary, some corrected positions still fail 3D projection because the face mask and the actual mesh surface do not perfectly overlap. When a border keypoint has no valid 3D position after mask-boundary correction, we search for the nearest pixel on the actual mesh surface and compute the 3D position directly from the mesh geometry at that pixel.
  \item \textbf{3D anchor-based fallback.} When a keypoint's 2D-to-3D projection fails (e.g., the 2D position lands in an area with no mesh geometry), we fall back to computing its 3D position from the already-resolved 3D positions of neighboring anchor keypoints, maintaining a running map of resolved positions so that later keypoints can reference those resolved earlier.
  \item \textbf{Adaptive eye mask selection.} Both segmentation-based and landmark-based eye masks can contribute to the face boundary computation, but including both can produce an overly large convex hull. We use only one eye mask source, selected adaptively: human assets use landmark-based eyes (higher precision), non-human assets use segmentation-based eyes (more robust), with auto-fallback if the preferred source is unavailable.
\end{enumerate}

After these correspondence fallbacks, we apply standard global similarity and regularized non-rigid registration.

\subsubsection{Template Fitting Optimization}

Given the template mesh and computed keypoint correspondences, the fitting stage deforms the template to match the input asset via a two-stage optimization.

\emph{Stage 1: Global Similarity Alignment.} We find the best similarity transformation (rotation $\mathbf{R}$, uniform scale $s$, translation $\mathbf{t}$) aligning template keypoints $\mathbf{v}_i$ to asset keypoints $\mathbf{p}_i$ by minimizing:
\begin{equation}
  E_{\text{sim}} = \sum_{i \in \mathcal{K}} \| s\mathbf{R}\mathbf{v}_i + \mathbf{t} - \mathbf{p}_i \|^2
\end{equation}
where $\mathcal{K}$ is the set of corresponding keypoints.

\emph{Stage 2: Per-Vertex Non-Rigid Deformation.} We optimize per-vertex displacements $\mathbf{d}_i$ for all vertices to minimize a composite energy function:
\begin{equation}
  E(\mathbf{D}) = \lambda_1 E_{\text{corr}} + \lambda_2 E_{\text{smooth}} + \lambda_3 E_{\text{edge}} + \lambda_4 E_{\text{tri}} + \lambda_5 E_{\text{flip}} + \lambda_6 E_{\text{reg}}
\end{equation}
where each energy term controls a different aspect of data fitting or geometric regularization to make this fitting pipeline robust to various input cases. We set $\lambda_1=20.0$ for the correspondence term and $\lambda_2,\ldots,\lambda_6=1.0$ for the geometric regularizers. The larger correspondence weight drives the sparse anchors toward their targets, while the smoothness, edge-length, triangle-shape, flip-prevention, and displacement terms preserve local structure in an elastic/ARAP-style manner. The terms are defined as follows:

\textbf{Correspondence energy} ($E_{\text{corr}}$) is the main data term that pulls the template keypoints toward their corresponding asset keypoints:
\begin{equation}
  E_{\text{corr}} = \sum_{i \in \mathcal{K}} \mathcal{H}(\| \mathbf{v}'_i - \mathbf{p}_i \|)
\end{equation}
where $\mathbf{v}'_i = \mathbf{v}_i + \mathbf{d}_i$ represents the deformed template vertex positions, $\mathbf{p}_i$ represents target landmark points, and $\mathcal{H}$ is a robust penalty function such as a Huber loss to ignore outliers.

\textbf{Smoothness energy} ($E_{\text{smooth}}$) enforces smooth deformations by encouraging neighboring vertices connected by a mesh edge $\mathcal{E}$ to move in a similar way:
\begin{equation}
  E_{\text{smooth}} = \sum_{(i,j) \in \mathcal{E}} \| \mathbf{d}_i - \mathbf{d}_j \|^2
\end{equation}

\textbf{Edge-length preservation energy} ($E_{\text{edge}}$) encourages the deformation to remain locally rigid by preserving original edge lengths up to global scale:
\begin{equation}
  E_{\text{edge}} = \sum_{(i,j) \in \mathcal{E}} \left( \| \mathbf{v}'_i - \mathbf{v}'_j \| - \| \mathbf{v}_i - \mathbf{v}_j \| \right)^2
\end{equation}

\textbf{Triangle shape preservation energy} ($E_{\text{tri}}$) preserves local triangle shapes more strongly than edge-length preservation alone and prevents skinny or sheared triangles:
\begin{equation}
  E_{\text{tri}} = \sum_{t \in \mathcal{T}} \| \mathbf{G}_t - \mathbf{G}'_t \|_F^2
\end{equation}
where $\mathbf{G}_t$ and $\mathbf{G}'_t$ denote triangle shape matrices encoding the edge lengths and angles before and after deformation, respectively.

\textbf{Triangle flip penalty} ($E_{\text{flip}}$) penalizes orientation reversal and reduces triangle flips and foldovers to prevent self-intersections:
\begin{equation}
  E_{\text{flip}} = \sum_{t \in \mathcal{T}} \max(0, -\mathbf{n}_t \cdot \mathbf{n}'_t)^3
\end{equation}
where $\mathbf{n}_t$ and $\mathbf{n}'_t$ denote triangle normals before and after deformation, respectively.

\textbf{Offset regularization energy} ($E_{\text{reg}}$) discourages unnecessarily large deformations and improves numerical stability:
\begin{equation}
  E_{\text{reg}} = \sum_i \| \mathbf{d}_i \|^2
\end{equation}

For normalized meshes, this optimization achieves \textbf{near-zero alignment error} ($<10^{-5}$). The nose-landmark-free template design and triangle flip penalty improve fitting robustness across diverse face proportions (compact for cats and elongated for dogs) without category-specific optimization rules; end-to-end failures are reported separately through Success Rate.

\subsection{Blendshape Rig Construction}
\label{sec:blendshape}

Given the fitted face mesh produced by Stage~1 (\Cref{sec:fitting}), Stage~2 constructs a production-ready FACS blendshape rig in three steps. The original asset is assumed to provide only the visible surface and source textures, with no pre-existing teeth, gums, tongue, or oral-cavity geometry. Stage~2 therefore consumes three additional inputs: the original static asset mesh (for surrounding geometry and source textures), the teeth template selected by the VLM in Stage~1 (\Cref{sec:riggability}), and a canonical FACS blendshape template.

\textbf{(1) Face Mesh Fusion.} The fitted template positions are first transferred to a high-quality animated topology mesh via UV correspondence, then reprojected onto the input character geometry. Eye regions are inset using virtual eyeball centers (computed via Procrustes alignment of eyelid vertices) to allow eyelid closure without self-intersection. The prepared template then replaces the face region on the original mesh: a 3D convex hull is computed for a subtractive Boolean (cut out the old face), the template border vertices are stitched to the remaining mesh boundary via closest-point welding, and small-edge collapse removes degenerate polygons (merge in the new face). This ensures proper radial topology flow around eyes and mouth for high-quality facial deformation.

\begin{figure}[tbp]
  \centering
  \includegraphics[width=0.75\linewidth]{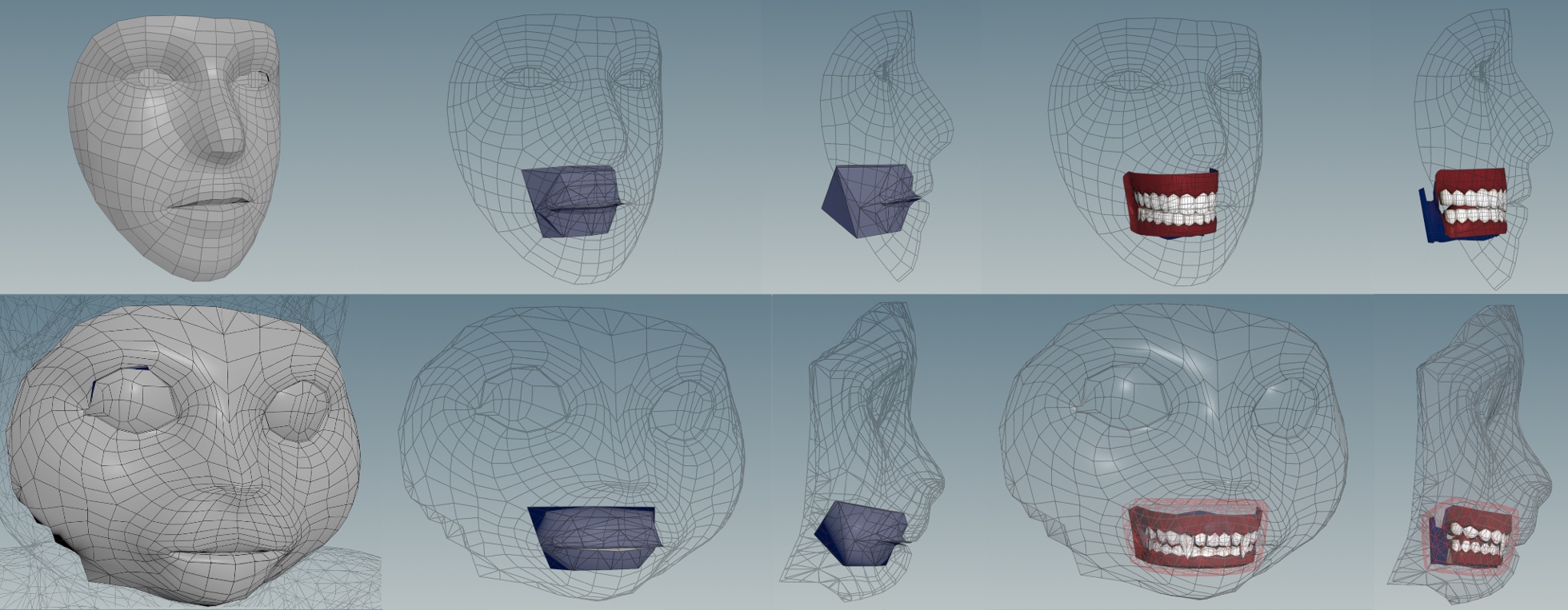}
  \caption{\textbf{Inner-mouth and teeth alignment.} Human (top) and animal (bottom) examples follow the same procedure. From left to right: the fitted face template, front and side views of the ARAP-deformed inner-mouth region, and front and side views of the aligned teeth archetype with its convex-hull collision proxy.}
  \label{fig:teeth_alignment}
\end{figure}

\textbf{(2) Teeth Registration and Texture Baking.} \emph{Teeth registration.} Because generated characters contain no oral-cavity geometry, the canonical face template specifies an inner-mouth region to which each teeth archetype is spatially registered at rest. After the character-specific face template is fitted to the target surface, we deform this inner-mouth region to the target shape using as-rigid-as-possible (ARAP) deformation~\citep{sorkine2007rigid}. We then estimate the best-fitting rigid transformation between the canonical and deformed inner-mouth regions using singular value decomposition (SVD) and apply it to the selected teeth archetype. A convex hull constructed from the aligned teeth serves as a collision proxy. At rest, an SDF-based refinement resolves initial collisions, which are particularly common for elongated dog snouts, by moving intersecting outer-face points out of the teeth volume. The selected archetype supplies the teeth, gums, and tongue as a consistent inner-mouth assembly (\Cref{fig:teeth_alignment}). \emph{Texture baking.} After mesh merging, UV islands are re-packed using an automated UV layout solver to maximize packing area with no overlap between shells, with sufficient padding to prevent visible seams during mipmap rendering. The face region is scaled up to allocate approximately 10$\times$ more texel density than non-face regions, ensuring high-quality facial texture across levels of detail (LOD). The original texture is resampled onto the new UV layout using a vectorized approach: for each pixel in the output texture, we find the barycentric coordinate and primitive ID on the merged mesh's UV layout, convert to world-space position, find the closest point on the original mesh, and sample the source texture at the corresponding UV. Teeth geometry uses a separate texture channel.

\begin{figure}[tbp]
  \centering
  \includegraphics[width=0.45\linewidth]{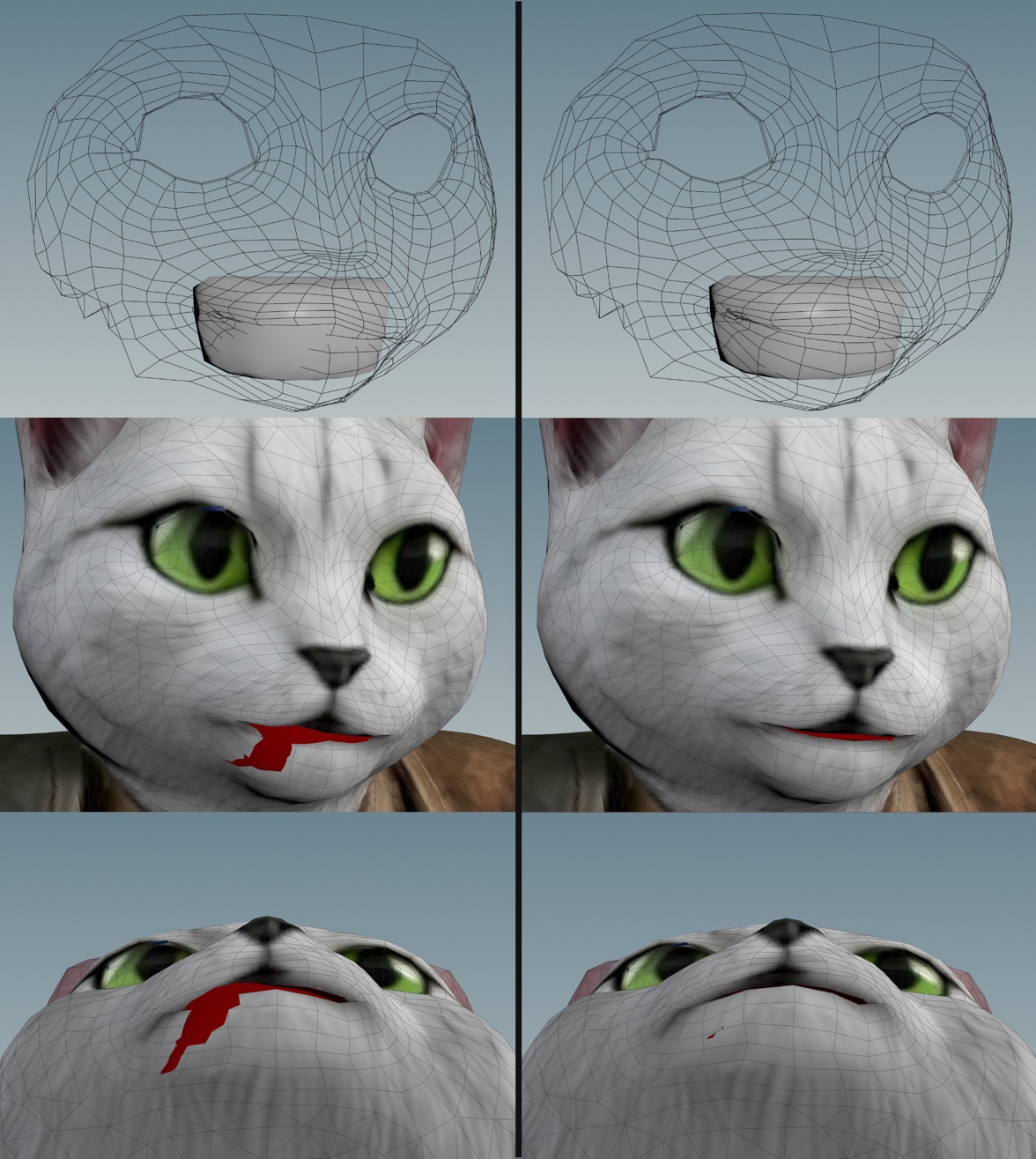}
  \caption{\textbf{Collision-aware blendshape transfer.} Standard deformation transfer (left) can move the inner-mouth geometry, shown in red, visibly through the outer facial surface during expressions such as AU12. Our collision-aware deformation transfer (right) uses the convex hull of the aligned teeth as a point-to-SDF collision constraint, substantially reducing visible intersections.}
  \label{fig:collision_transfer}
\end{figure}

\textbf{(3) FACS Blendshape Transfer.} Facial expressions are created by transferring blendshapes from the canonical FACS blendshape template to the merged character mesh using \textbf{sparse deformation transfer}~\citep{sumner2004deformation}. For each reference shape, triangle-level affine deformations are computed and transferred to the target mesh's triangle structure, preserving character-specific geometry while reproducing the expression. The transferred shapes are applied to the merged mesh using Point Deform with Delta Mush smoothing to blend into adjacent non-template regions. Post-processing includes: (a) \emph{contact-aware eyelid closure}~\citep{lewis2014practice}---upper eyelid vertices are snapped to lower lids and edge-relaxed for closed-eye shapes; (b) \emph{eye gaze shapes}---rotation around the virtual eyeball center; and (c) \emph{collision-aware deformation transfer}. Standard deformation transfer can move the outer face surface through the inner-mouth geometry, particularly for expressions with strong lip motion such as FACS Action Unit 12 (AU12, the lip-corner puller). We therefore reuse the convex hull of the aligned teeth as a collision surface during gradient-based shape transfer. The collision term is formulated as a point-to-SDF penalty: outer-face points are penalized when they enter the SDF of the teeth hull. This constraint is applied for every transferred expression, substantially reducing visible lip--inner-mouth intersections (\Cref{fig:collision_transfer}). Teeth geometry follows jaw-related shapes via radial basis function (RBF) warping converted to rigid transforms. The system supports three configurable shape-set tiers selected per application: \textbf{Core} (13 shapes) for basic dialog and emotion on mobile characters, \textbf{Additional} (46 shapes) for finer muscle control suitable for mid-fidelity game characters, and \textbf{Full} (155 shapes) for production VR applications where expression fidelity is paramount.

The output of Stage~2 is the final \textbf{FACS blendshape rig}, ready to be driven by arbitrary expression weights at run-time. Additional implementation details for face-region fusion, inner-mouth construction, UV/texture transfer, and expression post-processing are provided in Appendix~\ref{supp:sec:stage2}.

\section{Experiments}
\label{sec:experiments}

\subsection{Setup}

Rig-quality comparisons use screened Omni-Bench candidate subsets with independent artist-authored reference rigs (\Cref{subsec:metrics}); the riggability and segmentation evaluations use the sets described with \Cref{tab:riggability,tab:seg_ablation}. All inference-time experiments use a single NVIDIA A100 GPU; segmentation fine-tuning is an offline step on the multi-GPU setup below.

\subsection{Implementation Details}
\label{subsec:implementation}

\textbf{Sapiens Pre-Training.} We use Sapiens-1B~\citep{khirodkar2025sapiens} as the base segmentation backbone for all Sapiens-based parsers in OmniFaceRig. Following \Cref{sec:seg_models}, the encoder is initialized with masked-reconstruction pretraining: we first train at 512$\times$512 resolution for efficiency, then briefly switch to 1024$\times$1024 resolution for high-resolution facial boundary cues. The pretraining corpus contains approximately 4B curated realistic human images. We then perform a mid-stage unsupervised adaptation on 20M in-domain stylized character images. In our downstream parsing experiments, these two stages improve segmentation accuracy by 2--3\%, and the in-domain adaptation makes the subsequent supervised parser training more annotation-efficient.

\textbf{Fine-Tuned Sapiens Training.} To extend Sapiens' face parsing capability to stylized and non-human characters, we construct a custom training dataset of approximately 10,000 annotated 2D images covering humans, humanoids, and animal species such as felines, canines, ursines, and other stylized creatures, all rendered in frontal T-pose. We use an 80/10/10 train/validation/test split. The dataset contains approximately 60\% animal, 20\% stylized-humanoid, and 20\% human examples; this animal-heavy distribution targets the largest domain gap of the human-centric pretrained backbone. Starting from the pretrained encoder, we attach a lightweight three-stage deconvolutional decoder and train two compatible parser heads: a 41-class stylized parser for human-like characters and a 38-class humanoid parser for cartoon animals and strongly non-human characters. Both heads use the same curated image pool with task-specific label vocabularies. For evaluation and geometric fitting, these fine-grained labels are collapsed into the face-relevant cues consumed by the rigging pipeline: face boundary, eyes, nose or muzzle extent, mouth opening, teeth or inner-mouth support, and accessory-aware occlusion regions.

During fine-tuning, we freeze the Sapiens-1B ViT encoder to preserve its open-world visual features and train only the segmentation decoder heads. Training uses AdamW with a learning rate of $1 \times 10^{-4}$, cosine decay with 500 warmup steps, batch size 16, and class-balanced cross-entropy loss for imbalanced facial regions. Training runs for 50 epochs on 8$\times$ NVIDIA A100 GPUs, taking approximately 24 hours. We apply standard data augmentation: random horizontal flip, color jitter, random rotation ($\pm$15$^\circ$), and random scale (0.8--1.2$\times$). The fine-tuning images cannot be publicly released because the internally collected and manually annotated data are subject to licensing restrictions; we instead report the label configuration, model architecture, training protocol, and evaluation definitions.

The fine-tuned parsers achieve a mean Intersection-over-Union (mIoU) of \textbf{87.3\%} on our held-out test set after collapsing predictions to the rigging-relevant facial cues, compared to 62.1\% for the original Sapiens parser. The improvement is most pronounced on animal faces: cat face mIoU improves from 31.2\% to 84.5\%, and dog face mIoU from 28.7\% to 81.9\%, while human face mIoU remains stable (89.1\% $\rightarrow$ 91.2\%).

\begin{table}[!tbp]
  \centering
  \caption{\textbf{Human/humanoid evaluation against independent artist-authored references.} Errors use each method's valid outputs; Success Rate uses all 300 inputs. $^*$Requires manual correspondence annotations.}
  \label{tab:omnibench_comparison}
  \small
  \begin{NiceTabular}{l@{\hspace{4pt}}c@{\hspace{4pt}}c@{\hspace{4pt}}c@{\hspace{4pt}}c}
\CodeBefore
  \rowcolor{metabg}{5}
\Body
    \toprule
    Method & MAE$\downarrow$ & Q95$\downarrow$ & Penet.$\downarrow$ & Succ.$\uparrow$ \\
    & (mm) & (mm) & (\%) & (\%) \\
    \midrule
    DT~\citep{sumner2004deformation}$^*$ & 3.12 & 9.05 & -- & 82.0 \\
    NFR~\citep{qin2023neural} & 2.95 & 8.11 & -- & 85.3 \\
    \textbf{OmniFaceRig (Ours)} & \textbf{0.92} & \textbf{2.71} & \textbf{0.08} & \textbf{99.0} \\
    \bottomrule
  \end{NiceTabular}
\end{table}

\FloatBarrier
\begin{figure}[t]
  \centering
  \includegraphics[width=0.9\linewidth]{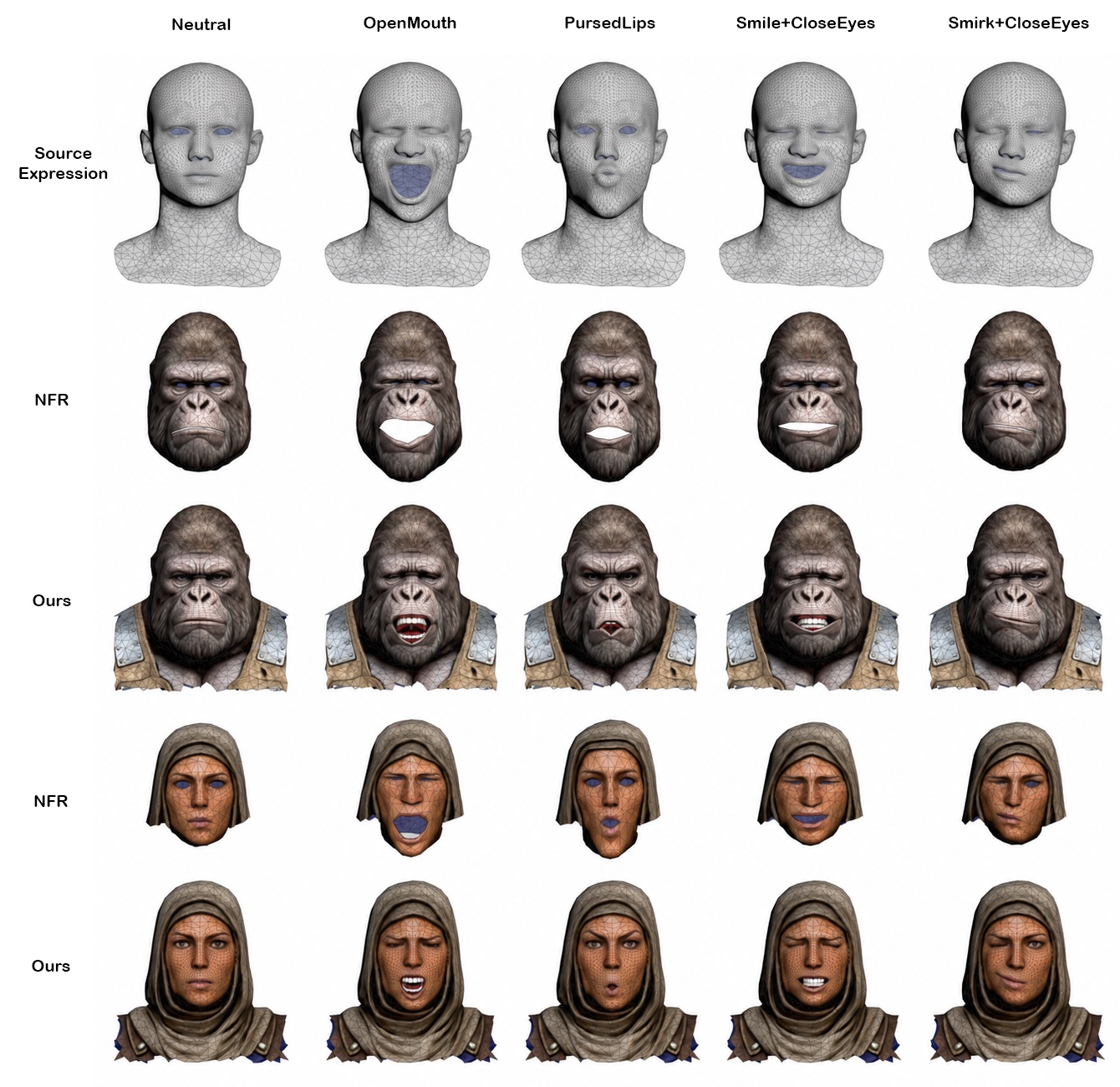}
  \caption{\textbf{Qualitative comparison with NFR.} Top: NFR source-template expressions (descriptive labels). Paired rows compare NFR and OmniFaceRig on a gorilla and a human; NFR uses preprocessed head-only targets, while OmniFaceRig poses are selected for correspondence.}
  \label{fig:nfr_qualitative}
\end{figure}

\subsection{Evaluation Metrics and Baselines}
\label{subsec:metrics}

Drawing upon the evaluation protocols established by recent state-of-the-art methods~\citep{ma2025riganyface, qin2023neural, cha2025neural}, we employ a comprehensive set of metrics. Before evaluation, each full-body T-pose mesh is isotropically normalized to fit a unit sphere in meter-scale coordinates, where one coordinate unit corresponds to one meter. We use Mean Absolute Error (MAE) and 95th-Percentile Vertex Error (Q95) to measure the mean and worst-case per-vertex Euclidean distance between the predicted expression mesh and ground truth, and report these distances in millimeters. Penetration Rate is the percentage of outer-face vertices lying inside the signed-distance field (SDF) of the aligned-teeth convex hull, aggregated over all expression--vertex tests of valid outputs. To quantify pipeline robustness, we introduce Success Rate, tracking the percentage of successfully rigged assets among screened candidate inputs that have passed the VLM+CV riggability checker. Finally, Processing Latency measures the end-to-end execution time in seconds.

All rig-quality evaluations (\Cref{tab:omnibench_comparison} and the animal results in the following subsection) use subsets of the screened Omni-Bench candidate inputs for which independently artist-authored reference rigs are available. We compare against Neural Face Rigging (NFR)~\citep{qin2023neural} and Deformation Transfer (DT)~\citep{sumner2004deformation} on the overlapping human/humanoid setting. We do not directly re-evaluate RigAnyFace~\citep{ma2025riganyface}, because neither its implementation nor the evaluation data required to reproduce its published numbers are publicly available.

\subsection{Quantitative Results}

\Cref{tab:omnibench_comparison} uses independent artist-authored ground truth, not corrected OmniFaceRig outputs. For the same 300 surface-only inputs, artists authored the target inner-mouth geometry and FACS blendshapes. The 200-human portion includes realistic/photorealistic faces; the remaining 100 are stylized humanoids. For a fair comparison, MAE and Q95 are computed only over corresponding outer-face vertices for all methods; the disconnected teeth, gums, and tongue components are excluded from both OmniFaceRig outputs and artist-authored references. Errors use each method's valid outputs, whereas Success Rate covers all 300 inputs: DT succeeds on 246, NFR on 256, and OmniFaceRig on 297. Baselines are restricted to this setting because DT and NFR do not produce equivalent animal rigs with generated inner mouths.

NFR's prescribed preparation requires manually isolating the head, removing garments, internal sockets, and eyeballs, cutting the eye and mouth openings, and normalizing the target. OmniFaceRig uses the original surface-only full-character mesh without per-asset preprocessing, better preserving the pursed lips (\Cref{fig:nfr_qualitative}, third column) and handling the non-human face.

\begin{figure}[p]
  \centering
  \includegraphics[width=\textwidth,height=0.88\textheight,keepaspectratio]{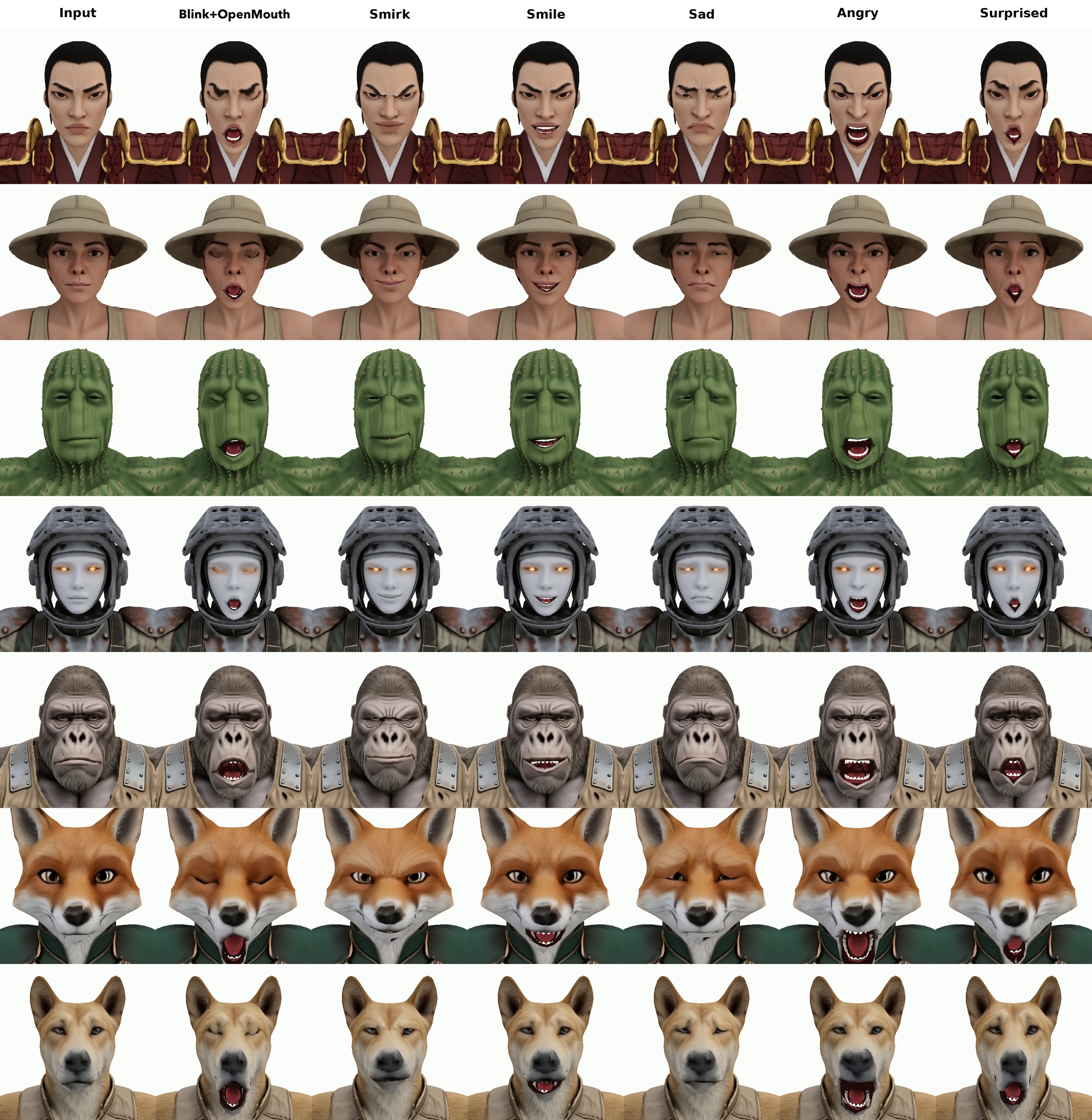}
  \caption{\textbf{Qualitative results across diverse Omni-Bench characters.} Each row shows a surface-only input (left) and six generated expressions with inner-mouth geometry. The same automatic pipeline supports humans, humanoids, and long- and short-muzzled animals while reducing teeth--face intersections through rest-pose and per-expression SDF refinement.}
  \label{fig:qualitative_results}
\end{figure}

On 200 animals with topology-compatible artist-authored references, OmniFaceRig produces 195 valid rigs (97.5\% success). Valid outputs obtain 2.60\,mm MAE, 6.77\,mm Q95, and 0.45\% penetration. \Cref{fig:qualitative_results} shows surface-only inputs alongside generated expressions and inner-mouth geometry across human, humanoid, and animal characters.

\begin{figure}[tbp]
  \centering
  \includegraphics[width=0.75\linewidth]{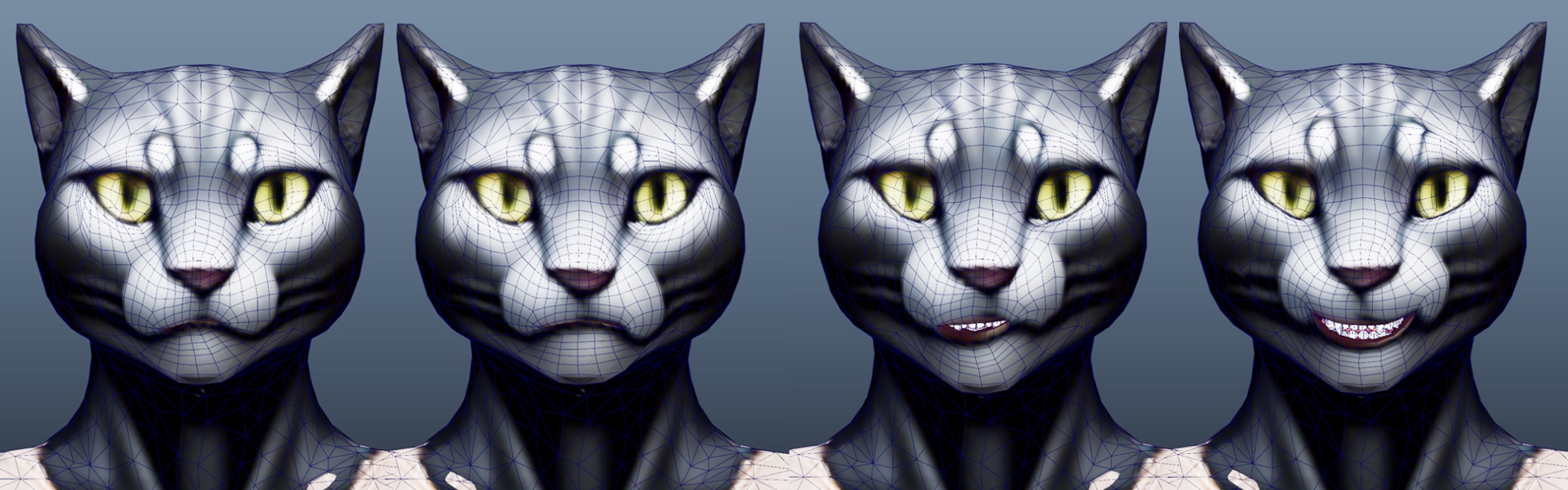}
  \caption{\textbf{Animal artist-reference comparison.} From left to right, the four columns show our result and the independently artist-authored reference for neutral, followed by the same pair for smile. The reference has more regular lower-lip topology and improved smile symmetry.}
  \label{fig:animal_reference}
\end{figure}

\begin{figure}[tbp]
  \centering
  \includegraphics[width=0.75\linewidth]{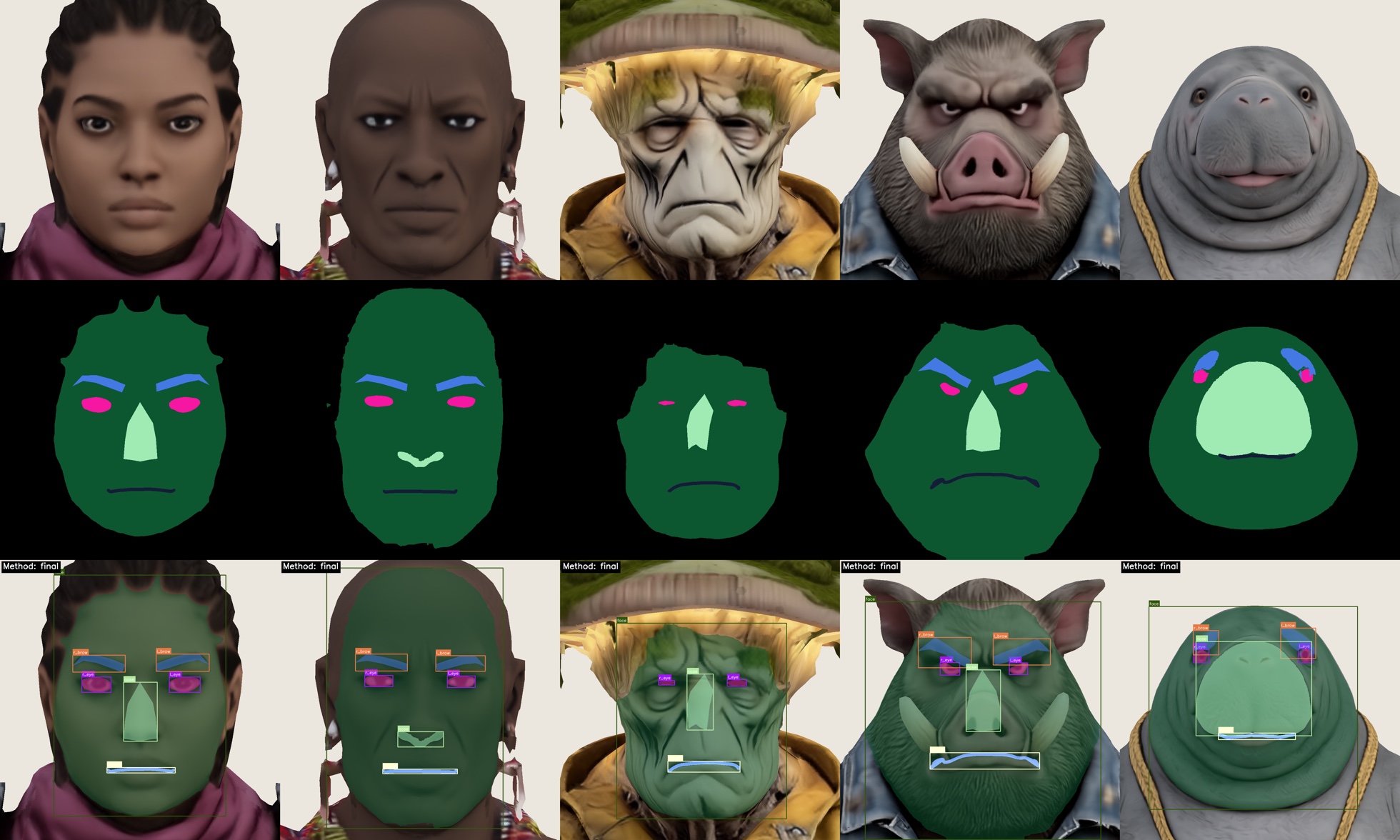}
  \caption{\textbf{Why nose anchors are unreliable.} Nose detection may fail or return inconsistent contours across human, humanoid, and animal faces. Rows show the input, mask, and overlay. This variation makes uniformly sampled nose-boundary points unreliable and motivates our nose-landmark-free design (\Cref{sec:fitting}).}
  \label{fig:nose_rationale}
\end{figure}

\begin{figure}[tbp]
  \centering
  \includegraphics[width=0.75\linewidth]{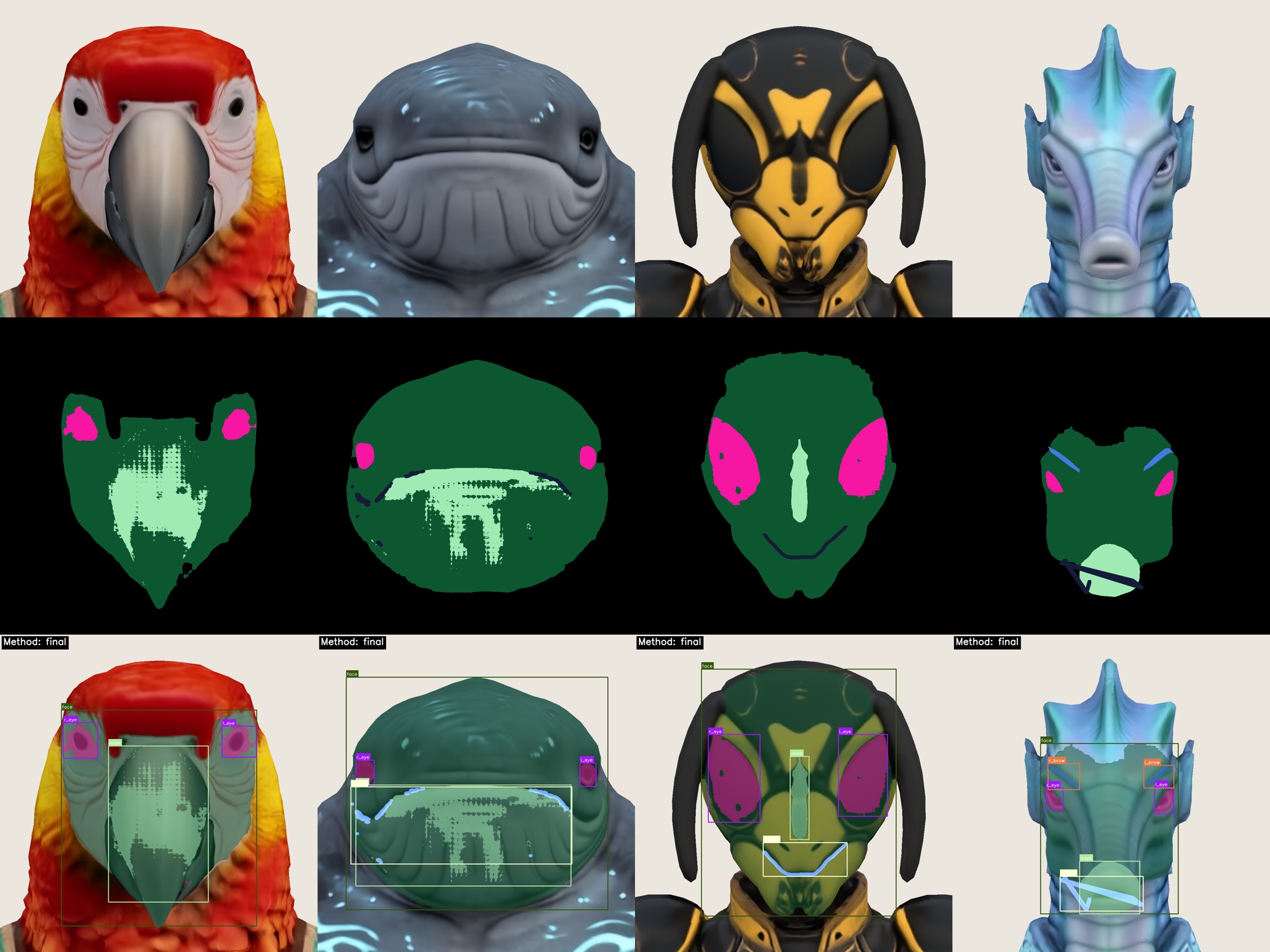}
  \caption{\textbf{Failure cases.} A sharp-beaked bird, wide-mouthed fish-like face, insect, and strongly elongated seahorse-like face challenge segmentation or fall outside our three template families. Rows show the input, mask, and overlay.}
  \label{fig:failures}
\end{figure}

\subsection{Ablation Studies}

\textbf{Segmentation Model Ablation.} We evaluate the contribution of each model in our four-model ensemble (\Cref{tab:seg_ablation}) by comparing face detection recall on the Omni-Bench evaluation inputs for individual models and the full ensemble.

\begin{table}[tbp]
  \centering
  \caption{Ablation: segmentation model contribution. Face detection recall (\%) on the Omni-Bench evaluation inputs. The four-model ensemble achieves nearly perfect recall by combining each model's strengths.}
  \label{tab:seg_ablation}
  \begin{NiceTabular}{lcccc}
\CodeBefore
  \rowcolor{metabg}{6}
\Body
    \toprule
    Model Config & Human & Cat & Dog & Overall \\
    \midrule
    Face landmark only & 92\% & 12\% & 18\% & 41\% \\
    Sapiens only & 88\% & 35\% & 42\% & 55\% \\
    SAM 3 only & 78\% & 71\% & 68\% & 72\% \\
    Fine-tuned Sapiens only & 90\% & 89\% & 85\% & 88\% \\
    \textbf{4-model ensemble (Ours)} & \textbf{100\%} & \textbf{100\%} & \textbf{97\%} & \textbf{$\sim$99\%} \\
    \bottomrule
  \end{NiceTabular}
\end{table}

Face landmark detection achieves high recall on human faces (92\%) but fails almost entirely on animals (12--18\%), as it is trained exclusively on human face data. Standard Sapiens degrades on non-human faces (35--42\%) due to its human-centric training, despite being a strong foundation model. SAM 3 offers the broadest single-model coverage (72\% overall) due to its class-agnostic design, but provides coarser segmentation boundaries than landmark-based methods. Fine-tuned Sapiens substantially improves animal face detection (85--89\%) while retaining strong human performance (90\%), validating the effectiveness of our fine-tuning strategy. The four-model ensemble with adaptive per-region selection achieves nearly 100\% overall recall by combining the accuracy of face landmark detection on humans, fine-tuned Sapiens' robustness on animals, and SAM 3's fallback coverage for edge cases.

\textbf{Nose Landmark Ablation.} We evaluate the impact of our nose-landmark-free template design on animal face rigging quality. When nose anchor keypoints are included (as in prior template designs), the pipeline produces severely distorted nose-mouth regions on 78\% of animal assets and crashes on 10\%, driven by the two failure modes discussed in \Cref{sec:fitting}: incorrect nose segmentation, and---even when the nose mask is plausible---unstable keypoints from uniform boundary sampling on small, irregular nose contours. Removing nose anchors and allowing the optimization to position the nose region from the surrounding border, eye, and mouth anchors eliminates these distortion artifacts and removes crashes attributable to nose-anchor instability in this ablation. \Cref{fig:nose_rationale} provides visual evidence for both failure modes across a range of character topologies.

\begin{samepage}
\textbf{Riggability Checker.} The hybrid VLM+CV approach outperforms either component alone (\Cref{tab:riggability}): the VLM provides strong archetype identification and accessory detection, while CV filtering catches geometric edge cases (no-face assets, degenerate meshes).
\par
\end{samepage}

\subsection{Latency Breakdown}
\label{sec:latency}

The face pipeline executes on a single NVIDIA A100 GPU. End-to-end latency is approximately 20--30\,s per asset, including data I/O. Stage~1 (riggability assessment, face parsing, keypoint extraction, and template fitting) takes 8--10\,s, while Stage~2 (mesh fusion, inner-mouth fitting, UV/texture processing, and blendshape transfer) takes 10--15\,s. Compared to manual sculpting workflows that typically take hours to days, this represents a reduction of 2--3 orders of magnitude.

\FloatBarrier
\section{Limitations and Future Work}
\label{sec:limitations}

OmniFaceRig is limited to visible-faced characters covered by our human, long-muzzle, and short-muzzle template families. Unusual anatomies---including sharp beaks, wide-mouthed fish-like faces, compound eyes, or strongly elongated snouts (\Cref{fig:failures})---may be rejected or fitted unreliably. Extending coverage requires additional templates, segmentation labels, and riggability criteria. Even a valid completed rig can retain localized fitting or penetration artifacts under extreme mouth motion or unusual muzzle geometry; \Cref{fig:animal_reference} compares one such near-failure with its artist-authored reference. The 0.45\% animal penetration rate and 97.5\% success rate quantify these remaining cases rather than implying perfectly intersection-free output. The 20--30\,s pipeline targets offline or batch processing rather than real-time editing. Future work will broaden topology coverage and further integrate perception with geometry. Additional riggability-checker failure modes (annotation borderline cases and VLM-only semantic errors) are discussed in Appendix~\ref{supp:sec:failure_modes}.

\section{Conclusion}
\label{sec:conclusion}

We presented OmniFaceRig, a fully automatic inner-mouth-aware face rigging pipeline for 3D characters. Given a static surface-only 3D mesh with no pre-modeled oral cavity, OmniFaceRig produces a production-ready rig with up to 155 FACS blendshapes and procedurally generated inner-mouth geometry (teeth, gums, and tongue), with no manual intervention at any stage. Our segmentation ensemble, topology-specific templates, and collision treatment support human and animal topologies; independent artist references show high success with low error and penetration. We also release Omni-Bench, to our knowledge the first open benchmark of more than 1,000 rigged biped characters with FACS facial blendshapes and inner-mouth geometry, to catalyze future research in automatic facial rigging.

\bibliographystyle{assets/plainnat}
\bibliography{references}

\begin{thebibliography}{74}
\providecommand{\natexlab}[1]{#1}
\providecommand{\url}[1]{\texttt{#1}}
\expandafter\ifx\csname urlstyle\endcsname\relax
  \providecommand{\doi}[1]{doi: #1}\else
  \providecommand{\doi}{doi: \begingroup \urlstyle{rm}\Url}\fi

\bibitem[Agrawal et~al.(2025)Agrawal, Akinyemi, Alvero, Behrooz, Buffalini,
  Carlucci, Chen, Chen, Chen, Cheng, Chowdary, Chuang, D'Avirro, Daly, Dong,
  Duppenthaler, Gao, Girard, Gleize, Gomez, Gong, Govindarajan, Han, He,
  Hernandez, Hristov, Huang, Inaguma, Jain, Janardhan, Jia, Klaiber, Kovachev,
  Kumar, Li, Li, Litvin, Liu, Ma, Ma, Ma, Ma, Mantovani, Miglani, Mohan,
  Morency, Ng, Ng, Nguyen, Oberai, Peloquin, Pino, Popovi{\'c}, Poursaeed,
  Prada, Rakotoarison, Ranjan, Richard, Ropers, Saleem, Sharma, Shcherbyna,
  Shen, Shen, Stathopoulos, Sun, Tomasello, Tran, Turkatenko, Wan, Wang, Wang,
  Williamson, Wood, Xiang, Yang, Yao, Zhang, Zhang, Zhang, Zheng, Zhyzheria,
  Zikes, and Zollhoefer]{agrawal2025seamless}
Vasu Agrawal, Akinniyi Akinyemi, Kathryn Alvero, Morteza Behrooz, Julia
  Buffalini, Fabio~Maria Carlucci, Joy Chen, Junming Chen, Zhang Chen, Shiyang
  Cheng, Praveen Chowdary, Joe Chuang, Antony D'Avirro, Jon Daly, Ning Dong,
  Mark Duppenthaler, Cynthia Gao, Jeff Girard, Martin Gleize, Sahir Gomez,
  Hongyu Gong, Srivathsan Govindarajan, Brandon Han, Sen He, Denise Hernandez,
  Yordan Hristov, Rongjie Huang, Hirofumi Inaguma, Somya Jain, Raj Janardhan,
  Qingyao Jia, Christopher Klaiber, Dejan Kovachev, Moneish Kumar, Hang Li,
  Yilei Li, Pavel Litvin, Wei Liu, Guangyao Ma, Jing Ma, Martin Ma, Xutai Ma,
  Lucas Mantovani, Sagar Miglani, Sreyas Mohan, Louis-Philippe Morency, Evonne
  Ng, Kam-Woh Ng, Tu~Anh Nguyen, Amia Oberai, Benjamin Peloquin, Juan Pino,
  Jovan Popovi{\'c}, Omid Poursaeed, Fabian Prada, Alice Rakotoarison, Rakesh
  Ranjan, Alexander Richard, Christophe Ropers, Safiyyah Saleem, Vasu Sharma,
  Alex Shcherbyna, Jia Shen, Jie Shen, Anastasis Stathopoulos, Anna Sun, Paden
  Tomasello, Tuan Tran, Arina Turkatenko, Bo~Wan, Chao Wang, Jeff Wang, Mary
  Williamson, Carleigh Wood, Tao Xiang, Yilin Yang, Julien Yao, Chen Zhang,
  Jiemin Zhang, Xinyue Zhang, Jason Zheng, Pavlo Zhyzheria, Jan Zikes, and
  Michael Zollhoefer.
\newblock {Seamless Interaction}: Dyadic audiovisual motion modeling and
  large-scale dataset.
\newblock \emph{arXiv preprint arXiv:2506.22554}, 2025.

\bibitem[{Blanco i Ribera} et~al.(2017){Blanco i Ribera}, Zell, Lewis, Noh, and
  Botsch]{ribera2017facial}
Roger {Blanco i Ribera}, Eduard Zell, J.~P. Lewis, Junyong Noh, and Mario
  Botsch.
\newblock Facial retargeting with automatic range of motion alignment.
\newblock \emph{ACM Transactions on Graphics}, 36\penalty0 (4):\penalty0
  154:1--154:12, 2017.
\newblock \doi{10.1145/3072959.3073674}.

\bibitem[Blanz and Vetter(1999)]{blanz1999morphable}
Volker Blanz and Thomas Vetter.
\newblock A morphable model for the synthesis of {3D} faces.
\newblock In \emph{Proceedings of the 26th Annual Conference on Computer
  Graphics and Interactive Techniques (SIGGRAPH)}, pages 187--194, 1999.
\newblock \doi{10.1145/311535.311556}.

\bibitem[Brunton et~al.(2014)Brunton, Bolkart, and
  Wuhrer]{brunton2014multilinear}
Alan Brunton, Timo Bolkart, and Stefanie Wuhrer.
\newblock Multilinear wavelets: A statistical shape space for human faces.
\newblock In \emph{European Conference on Computer Vision (ECCV)}, pages
  297--312, 2014.
\newblock \doi{10.1007/978-3-319-10590-1_20}.

\bibitem[Cao et~al.(2014)Cao, Weng, Zhou, Tong, and Zhou]{cao2014facewarehouse}
Chen Cao, Yanlin Weng, Shun Zhou, Yiying Tong, and Kun Zhou.
\newblock {FaceWarehouse}: A {3D} facial expression database for visual
  computing.
\newblock \emph{IEEE Transactions on Visualization and Computer Graphics},
  20\penalty0 (3):\penalty0 413--425, 2014.
\newblock \doi{10.1109/TVCG.2013.249}.

\bibitem[Carion et~al.(2026)Carion, Gustafson, Hu, Debnath, Hu, {Suris
  Coll-Vinent}, Ryali, Alwala, Khedr, Huang, Lei, Ma, Guo, Kalla, Marks, Greer,
  Wang, Sun, R{\"a}dle, Afouras, Mavroudi, Xu, Wu, Zhou, Momeni, Hazra, Ding,
  Vaze, Porcher, Li, Li, Kamath, Cheng, Doll{\'a}r, Ravi, Saenko, Zhang, and
  Feichtenhofer]{carion2025sam3}
Nicolas Carion, Laura Gustafson, Yuan-Ting Hu, Shoubhik Debnath, Ronghang Hu,
  Didac {Suris Coll-Vinent}, Chaitanya Ryali, Kalyan~Vasudev Alwala, Haitham
  Khedr, Andrew Huang, Jie Lei, Tengyu Ma, Baishan Guo, Arpit Kalla, Markus
  Marks, Joseph Greer, Meng Wang, Peize Sun, Roman R{\"a}dle, Triantafyllos
  Afouras, Effrosyni Mavroudi, Katherine Xu, Tsung-Han Wu, Yu~Zhou, Liliane
  Momeni, Rishi Hazra, Shuangrui Ding, Sagar Vaze, Francois Porcher, Feng Li,
  Siyuan Li, Aishwarya Kamath, Ho~Kei Cheng, Piotr Doll{\'a}r, Nikhila Ravi,
  Kate Saenko, Pengchuan Zhang, and Christoph Feichtenhofer.
\newblock {SAM 3}: {Segment Anything} with concepts.
\newblock In \emph{International Conference on Learning Representations
  (ICLR)}, pages 138846--138923, 2026.

\bibitem[Cetinaslan and Orvalho(2020{\natexlab{a}})]{cetinaslan2020sketching}
Ozan Cetinaslan and Ver{\'o}nica Orvalho.
\newblock Sketching manipulators for localized blendshape editing.
\newblock \emph{Graphical Models}, 108:\penalty0 101059, 2020{\natexlab{a}}.
\newblock \doi{10.1016/j.gmod.2020.101059}.

\bibitem[Cetinaslan and Orvalho(2020{\natexlab{b}})]{cetinaslan2020stabilized}
Ozan Cetinaslan and Ver{\'o}nica Orvalho.
\newblock Stabilized blendshape editing using localized {J}acobian transpose
  descent.
\newblock \emph{Graphical Models}, 112:\penalty0 101091, 2020{\natexlab{b}}.
\newblock \doi{10.1016/j.gmod.2020.101091}.

\bibitem[Cha et~al.(2025)Cha, Yoon, Seo, and Noh]{cha2025neural}
Sihun Cha, Serin Yoon, Kwanggyoon Seo, and Junyong Noh.
\newblock Neural face skinning for mesh-agnostic facial expression cloning.
\newblock \emph{Computer Graphics Forum}, 44\penalty0 (2):\penalty0 e70009,
  2025.
\newblock \doi{10.1111/cgf.70009}.

\bibitem[Chandran et~al.(2020)Chandran, Bradley, Gross, and
  Beeler]{chandran2020semantic}
Prashanth Chandran, Derek Bradley, Markus Gross, and Thabo Beeler.
\newblock Semantic deep face models.
\newblock In \emph{International Conference on 3D Vision (3DV)}, pages
  345--354, 2020.
\newblock \doi{10.1109/3DV50981.2020.00044}.

\bibitem[Chandran et~al.(2022{\natexlab{a}})Chandran, Ciccone, Gross, and
  Bradley]{chandran2022local}
Prashanth Chandran, Lo{\"\i}c Ciccone, Markus Gross, and Derek Bradley.
\newblock Local anatomically-constrained facial performance retargeting.
\newblock \emph{ACM Transactions on Graphics}, 41\penalty0 (4):\penalty0
  168:1--168:14, 2022{\natexlab{a}}.
\newblock \doi{10.1145/3528223.3530114}.

\bibitem[Chandran et~al.(2022{\natexlab{b}})Chandran, Zoss, Gross, Gotardo, and
  Bradley]{chandran2022shape}
Prashanth Chandran, Gaspard Zoss, Markus Gross, Paulo F.~U. Gotardo, and Derek
  Bradley.
\newblock {Shape Transformers}: Topology-independent {3D} shape models using
  transformers.
\newblock \emph{Computer Graphics Forum}, 41\penalty0 (2):\penalty0 195--207,
  2022{\natexlab{b}}.
\newblock \doi{10.1111/cgf.14468}.

\bibitem[Chandran et~al.(2025)Chandran, Ciccone, Zoss, and
  Bradley]{chandran2025neural}
Prashanth Chandran, Lo{\"\i}c Ciccone, Gaspard Zoss, and Derek Bradley.
\newblock Neural facial deformation transfer.
\newblock In \emph{Eurographics 2025 -- Short Papers}. The Eurographics
  Association, 2025.
\newblock \doi{10.2312/egs.20251036}.

\bibitem[Cheng et~al.(2024)Cheng, Ma, and Pan]{cheng2024stylizedfacepoint}
Shengran Cheng, Chuhang Ma, and Ye~Pan.
\newblock {StylizedFacePoint}: Facial landmark detection for stylized
  characters.
\newblock In \emph{ACM International Conference on Multimedia (ACM MM)}, pages
  8072--8080, 2024.
\newblock \doi{10.1145/3664647.3680984}.

\bibitem[Cudeiro et~al.(2019)Cudeiro, Bolkart, Laidlaw, Ranjan, and
  Black]{cudeiro2019capture}
Daniel Cudeiro, Timo Bolkart, Cassidy Laidlaw, Anurag Ranjan, and Michael~J.
  Black.
\newblock Capture, learning, and synthesis of {3D} speaking styles.
\newblock In \emph{IEEE/CVF Conference on Computer Vision and Pattern
  Recognition (CVPR)}, pages 10093--10103, 2019.
\newblock \doi{10.1109/CVPR.2019.01034}.

\bibitem[Danieau et~al.(2019)Danieau, Gubins, Olivier, Dumas, Denis, Lopez,
  Mollet, Frager, and Avril]{danieau2019automatic}
Fabien Danieau, Ilja Gubins, Nicolas Olivier, Olivier Dumas, Bernard Denis,
  Thomas Lopez, Nicolas Mollet, Brian Frager, and Quentin Avril.
\newblock Automatic generation and stylization of {3D} facial rigs.
\newblock In \emph{IEEE Conference on Virtual Reality and 3D User Interfaces
  (VR)}, pages 784--792, 2019.
\newblock \doi{10.1109/VR.2019.8798208}.

\bibitem[Deng et~al.(2025)Deng, Zhang, Geng, Wu, and Wu]{anymate2025}
Yufan Deng, Yuhao Zhang, Chen Geng, Shangzhe Wu, and Jiajun Wu.
\newblock {Anymate}: A dataset and baselines for learning {3D} object rigging.
\newblock In \emph{ACM SIGGRAPH Conference Papers}, pages 112:1--112:10, 2025.
\newblock \doi{10.1145/3721238.3730743}.

\bibitem[Ekman and Friesen(1978)]{ekman1978facial}
Paul Ekman and Wallace~V. Friesen.
\newblock \emph{Facial Action Coding System: A Technique for the Measurement of
  Facial Movement}.
\newblock Consulting Psychologists Press, Palo Alto, CA, 1978.

\bibitem[Feng et~al.(2021)Feng, Feng, Black, and Bolkart]{feng2021learning}
Yao Feng, Haiwen Feng, Michael~J. Black, and Timo Bolkart.
\newblock Learning an animatable detailed {3D} face model from in-the-wild
  images.
\newblock \emph{ACM Transactions on Graphics}, 40\penalty0 (4):\penalty0
  88:1--88:13, 2021.
\newblock \doi{10.1145/3450626.3459936}.

\bibitem[{Fernandez Abrevaya} et~al.(2019){Fernandez Abrevaya}, Boukhayma,
  Wuhrer, and Boyer]{abrevaya2019decoupled}
Victoria {Fernandez Abrevaya}, Adnane Boukhayma, Stefanie Wuhrer, and Edmond
  Boyer.
\newblock A decoupled {3D} facial shape model by adversarial training.
\newblock In \emph{IEEE/CVF International Conference on Computer Vision
  (ICCV)}, pages 9418--9427, 2019.
\newblock \doi{10.1109/ICCV.2019.00951}.

\bibitem[Hanocka et~al.(2019)Hanocka, Hertz, Fish, Giryes, Fleishman, and
  Cohen-Or]{hanocka2019meshcnn}
Rana Hanocka, Amir Hertz, Noa Fish, Raja Giryes, Shachar Fleishman, and Daniel
  Cohen-Or.
\newblock {MeshCNN}: A network with an edge.
\newblock \emph{ACM Transactions on Graphics}, 38\penalty0 (4):\penalty0
  90:1--90:12, 2019.
\newblock \doi{10.1145/3306346.3322959}.

\bibitem[Jang et~al.(2025)Jang, Hong, Cha, and Kim]{jang2025controlface}
Wooseok Jang, Youngjun Hong, Geonho Cha, and Seungryong Kim.
\newblock {ControlFace}: Harnessing facial parametric control for face rigging.
\newblock In \emph{IEEE/CVF Conference on Computer Vision and Pattern
  Recognition (CVPR)}, pages 5614--5624, 2025.
\newblock \doi{10.1109/CVPR52734.2025.00527}.

\bibitem[Jiang et~al.(2019)Jiang, Wu, Chen, and Zhang]{jiang2019disentangled}
Zi-Hang Jiang, Qianyi Wu, Keyu Chen, and Juyong Zhang.
\newblock Disentangled representation learning for {3D} face shape.
\newblock In \emph{IEEE/CVF Conference on Computer Vision and Pattern
  Recognition (CVPR)}, pages 11949--11958, 2019.
\newblock \doi{10.1109/CVPR.2019.01223}.

\bibitem[Jun and Nichol(2023)]{jun2023shape}
Heewoo Jun and Alex Nichol.
\newblock {Shap-E}: Generating conditional {3D} implicit functions.
\newblock \emph{arXiv preprint arXiv:2305.02463}, 2023.

\bibitem[Kavan et~al.(2024)Kavan, Doublestein, Prazak, Cioffi, and
  Roble]{kavan2024compressed}
Ladislav Kavan, John Doublestein, Martin Prazak, Matthew Cioffi, and Doug
  Roble.
\newblock Compressed skinning for facial blendshapes.
\newblock In \emph{ACM SIGGRAPH Conference Papers}, pages 47:1--47:9, 2024.
\newblock \doi{10.1145/3641519.3657477}.

\bibitem[Khirodkar et~al.(2024)Khirodkar, Bagautdinov, Martinez, Su, James,
  Selednik, Anderson, and Saito]{khirodkar2025sapiens}
Rawal Khirodkar, Timur Bagautdinov, Julieta Martinez, Zhaoen Su, Austin James,
  Peter Selednik, Stuart Anderson, and Shunsuke Saito.
\newblock {Sapiens}: Foundation for human vision models.
\newblock In \emph{European Conference on Computer Vision (ECCV)}, volume 15062
  of \emph{Lecture Notes in Computer Science}, pages 206--228. Springer, 2024.
\newblock \doi{10.1007/978-3-031-73235-5_12}.

\bibitem[Kirillov et~al.(2023)Kirillov, Mintun, Ravi, Mao, Rolland, Gustafson,
  Xiao, Whitehead, Berg, Lo, Doll{\'a}r, and Girshick]{kirillov2023segment}
Alexander Kirillov, Eric Mintun, Nikhila Ravi, Hanzi Mao, Chloe Rolland, Laura
  Gustafson, Tete Xiao, Spencer Whitehead, Alexander~C. Berg, Wan-Yen Lo, Piotr
  Doll{\'a}r, and Ross Girshick.
\newblock {Segment Anything}.
\newblock In \emph{IEEE/CVF International Conference on Computer Vision
  (ICCV)}, pages 3992--4003, 2023.
\newblock \doi{10.1109/ICCV51070.2023.00371}.

\bibitem[Lewis and Anjyo(2010)]{lewis2010direct}
J.~P. Lewis and Ken-ichi Anjyo.
\newblock Direct manipulation blendshapes.
\newblock \emph{IEEE Computer Graphics and Applications}, 30\penalty0
  (4):\penalty0 42--50, 2010.
\newblock \doi{10.1109/MCG.2010.41}.

\bibitem[Lewis et~al.(2014)Lewis, Anjyo, Rhee, Zhang, Pighin, and
  Deng]{lewis2014practice}
J.~P. Lewis, Ken Anjyo, Taehyun Rhee, Mengjie Zhang, Frederic~H. Pighin, and
  Zhigang Deng.
\newblock Practice and theory of blendshape facial models.
\newblock In \emph{Eurographics 2014 -- State of the Art Reports}, pages
  199--218. The Eurographics Association, 2014.
\newblock \doi{10.2312/egst.20141042}.

\bibitem[Li et~al.(2010)Li, Weise, and Pauly]{li2010example}
Hao Li, Thibaut Weise, and Mark Pauly.
\newblock Example-based facial rigging.
\newblock \emph{ACM Transactions on Graphics}, 29\penalty0 (4):\penalty0
  32:1--32:6, 2010.
\newblock \doi{10.1145/1778765.1778769}.

\bibitem[Li et~al.(2020{\natexlab{a}})Li, Kuang, Zhao, He, Bladin, and
  Li]{li2020dynamic}
Jiaman Li, Zhengfei Kuang, Yajie Zhao, Mingming He, Karl Bladin, and Hao Li.
\newblock Dynamic facial asset and rig generation from a single scan.
\newblock \emph{ACM Transactions on Graphics}, 39\penalty0 (6):\penalty0
  215:1--215:18, 2020{\natexlab{a}}.
\newblock \doi{10.1145/3414685.3417817}.

\bibitem[Li et~al.(2020{\natexlab{b}})Li, Bladin, Zhao, Chinara, Ingraham,
  Xiang, Ren, Prasad, Kishore, Xing, and Li]{li2020learning}
Ruilong Li, Karl Bladin, Yajie Zhao, Chinmay Chinara, Owen Ingraham, Pengda
  Xiang, Xinglei Ren, Pratusha Prasad, Bipin Kishore, Jun Xing, and Hao Li.
\newblock Learning formation of physically-based face attributes.
\newblock In \emph{IEEE/CVF Conference on Computer Vision and Pattern
  Recognition (CVPR)}, pages 3407--3416, 2020{\natexlab{b}}.
\newblock \doi{10.1109/CVPR42600.2020.00347}.

\bibitem[Li et~al.(2017)Li, Bolkart, Black, Li, and Romero]{li2017learning}
Tianye Li, Timo Bolkart, Michael~J. Black, Hao Li, and Javier Romero.
\newblock Learning a model of facial shape and expression from {4D} scans.
\newblock \emph{ACM Transactions on Graphics}, 36\penalty0 (6):\penalty0
  194:1--194:17, 2017.
\newblock \doi{10.1145/3130800.3130813}.

\bibitem[Lin et~al.(2023)Lin, Gao, Tang, Takikawa, Zeng, Huang, Kreis, Fidler,
  Liu, and Lin]{lin2023magic3d}
Chen-Hsuan Lin, Jun Gao, Luming Tang, Towaki Takikawa, Xiaohui Zeng, Xun Huang,
  Karsten Kreis, Sanja Fidler, Ming-Yu Liu, and Tsung-Yi Lin.
\newblock {Magic3D}: High-resolution text-to-{3D} content creation.
\newblock In \emph{IEEE/CVF Conference on Computer Vision and Pattern
  Recognition (CVPR)}, pages 300--309, 2023.
\newblock \doi{10.1109/CVPR52729.2023.00037}.

\bibitem[Lin et~al.(2021)Lin, Shen, Wang, and Pantic]{lin2021roi}
Yiming Lin, Jie Shen, Yujiang Wang, and Maja Pantic.
\newblock {RoI Tanh-polar} transformer network for face parsing in the wild.
\newblock \emph{Image and Vision Computing}, 112:\penalty0 104190, 2021.
\newblock \doi{10.1016/j.imavis.2021.104190}.

\bibitem[Lin et~al.(2025)Lin, Shen, Wang, and Pantic]{lin2025fpage}
Yiming Lin, Jie Shen, Yujiang Wang, and Maja Pantic.
\newblock {FP-Age}: Leveraging face parsing attention for facial age estimation
  in the wild.
\newblock \emph{IEEE Transactions on Image Processing}, 34:\penalty0
  4767--4777, 2025.
\newblock \doi{10.1109/TIP.2022.3155944}.

\bibitem[Liu et~al.(2011)Liu, Ma, Chang, Wang, and Debevec]{liu2011framework}
Ko-Yun Liu, Wan-Chun Ma, Chun-Fa Chang, Chuan-Chang Wang, and Paul Debevec.
\newblock A framework for locally retargeting and rendering facial performance.
\newblock \emph{Computer Animation and Virtual Worlds}, 22\penalty0
  (2--3):\penalty0 159--167, 2011.
\newblock \doi{10.1002/cav.404}.

\bibitem[Liu et~al.(2023)Liu, Wu, Van~Hoorick, Tokmakov, Zakharov, and
  Vondrick]{liu2023zero123}
Ruoshi Liu, Rundi Wu, Basile Van~Hoorick, Pavel Tokmakov, Sergey Zakharov, and
  Carl Vondrick.
\newblock {Zero-1-to-3}: Zero-shot one image to {3D} object.
\newblock In \emph{IEEE/CVF International Conference on Computer Vision
  (ICCV)}, pages 9264--9275, 2023.
\newblock \doi{10.1109/ICCV51070.2023.00853}.

\bibitem[Lugaresi et~al.(2019)Lugaresi, Tang, Nash, McClanahan, Uboweja, Hays,
  Zhang, Chang, Yong, Lee, Chang, Hua, Georg, and
  Grundmann]{lugaresi2019mediapipe}
Camillo Lugaresi, Jiuqiang Tang, Hadon Nash, Chris McClanahan, Esha Uboweja,
  Michael Hays, Fan Zhang, Chuo-Ling Chang, Ming~Guang Yong, Juhyun Lee,
  Wan-Teh Chang, Wei Hua, Manfred Georg, and Matthias Grundmann.
\newblock {MediaPipe}: A framework for perceiving and processing reality.
\newblock In \emph{Third Workshop on Computer Vision for AR/VR at IEEE/CVF
  Conference on Computer Vision and Pattern Recognition (CVPR)}, 2019.

\bibitem[Luo et~al.(2023)Luo, Cai, Dong, Ming, Qiu, Zhan, and
  Han]{luo2023rabit}
Zhongjin Luo, Shengcai Cai, Jinguo Dong, Ruibo Ming, Liangdong Qiu, Xiaohang
  Zhan, and Xiaoguang Han.
\newblock {RaBit}: Parametric modeling of {3D} biped cartoon characters with a
  topological-consistent dataset.
\newblock In \emph{IEEE/CVF Conference on Computer Vision and Pattern
  Recognition (CVPR)}, pages 12825--12835, 2023.
\newblock \doi{10.1109/CVPR52729.2023.01233}.

\bibitem[Ma et~al.(2025)Ma, Kneubuehler, Chu, Sachs, Jiang, and
  Huang]{ma2025riganyface}
Wenchao Ma, Dario Kneubuehler, Maurice Chu, Ian Sachs, Haomiao Jiang, and
  Sharon~Xiaolei Huang.
\newblock {RigAnyFace}: Scaling neural facial mesh auto-rigging with unlabeled
  data.
\newblock In \emph{Advances in Neural Information Processing Systems
  (NeurIPS)}, volume~38, pages 51772--51796, 2025.
\newblock \doi{10.52202/085713-1543}.

\bibitem[Ming et~al.(2024)Ming, Li, Ling, Zhang, and Xu]{ming2024high}
Xin Ming, Jiawei Li, Jingwang Ling, Libo Zhang, and Feng Xu.
\newblock High-quality mesh blendshape generation from face videos via neural
  inverse rendering.
\newblock In \emph{European Conference on Computer Vision (ECCV)}, pages
  106--125, 2024.
\newblock \doi{10.1007/978-3-031-72897-6_7}.

\bibitem[Mohammadi et~al.(2026)Mohammadi, Weiss, Buhmann, Ciccone, Sumner,
  Bradley, and Guay]{canrig2026}
Arad Mohammadi, Sebastian Weiss, Jakob Buhmann, Lo{\"\i}c Ciccone, Robert~W.
  Sumner, Derek Bradley, and Martin Guay.
\newblock {CANRig}: Cross-attention neural face rigging with variable local
  control.
\newblock \emph{Computer Graphics Forum}, page e70426, 2026.
\newblock \doi{10.1111/cgf.70426}.

\bibitem[Noh and Neumann(2001)]{noh2001expression}
Jun-yong Noh and Ulrich Neumann.
\newblock Expression cloning.
\newblock In \emph{Proceedings of the 28th Annual Conference on Computer
  Graphics and Interactive Techniques (SIGGRAPH)}, pages 277--288, 2001.
\newblock \doi{10.1145/383259.383290}.

\bibitem[Orvalho et~al.(2012)Orvalho, Bastos, Parke, Oliveira, and
  Alvarez]{orvalho2012facial}
Ver{\'o}nica Orvalho, Pedro Bastos, Frederic~I. Parke, Bruno Oliveira, and
  Xenxo Alvarez.
\newblock A facial rigging survey.
\newblock In \emph{Eurographics 2012 -- State of the Art Reports}, pages
  183--204. The Eurographics Association, 2012.
\newblock \doi{10.2312/conf/EG2012/stars/183-204}.

\bibitem[Orvalho et~al.(2008)Orvalho, Zacur, and
  Susin]{orvalho2008transferring}
Ver{\'o}nica~Costa Orvalho, Ernesto Zacur, and Antonio Susin.
\newblock Transferring the rig and animations from a character to different
  face models.
\newblock \emph{Computer Graphics Forum}, 27\penalty0 (8):\penalty0 1997--2012,
  2008.
\newblock \doi{10.1111/j.1467-8659.2008.01187.x}.

\bibitem[Paysan et~al.(2009)Paysan, Knothe, Amberg, Romdhani, and
  Vetter]{paysan20093d}
Pascal Paysan, Reinhard Knothe, Brian Amberg, Sami Romdhani, and Thomas Vetter.
\newblock A {3D} face model for pose and illumination invariant face
  recognition.
\newblock In \emph{IEEE International Conference on Advanced Video and Signal
  Based Surveillance (AVSS)}, pages 296--301, 2009.
\newblock \doi{10.1109/AVSS.2009.58}.

\bibitem[Poole et~al.(2023)Poole, Jain, Barron, and
  Mildenhall]{poole2023dreamfusion}
Ben Poole, Ajay Jain, Jonathan~T. Barron, and Ben Mildenhall.
\newblock {DreamFusion}: Text-to-{3D} using {2D} diffusion.
\newblock In \emph{International Conference on Learning Representations
  (ICLR)}, 2023.

\bibitem[Qin et~al.(2023)Qin, Saito, Aigerman, Groueix, and
  Komura]{qin2023neural}
Dafei Qin, Jun Saito, Noam Aigerman, Thibault Groueix, and Taku Komura.
\newblock Neural face rigging for animating and retargeting facial meshes in
  the wild.
\newblock In \emph{ACM SIGGRAPH Conference Proceedings}, pages 68:1--68:11,
  2023.
\newblock \doi{10.1145/3588432.3591556}.

\bibitem[Ranjan et~al.(2018)Ranjan, Bolkart, Sanyal, and
  Black]{ranjan2018generating}
Anurag Ranjan, Timo Bolkart, Soubhik Sanyal, and Michael~J. Black.
\newblock Generating {3D} faces using convolutional mesh autoencoders.
\newblock In \emph{European Conference on Computer Vision (ECCV)}, Lecture
  Notes in Computer Science, pages 725--741. Springer, 2018.
\newblock \doi{10.1007/978-3-030-01219-9_43}.

\bibitem[Ravi et~al.(2025)Ravi, Gabeur, Hu, Hu, Ryali, Ma, Khedr, R{\"a}dle,
  Rolland, Gustafson, Mintun, Pan, Alwala, Carion, Wu, Girshick, Doll{\'a}r,
  and Feichtenhofer]{ravi2024sam2}
Nikhila Ravi, Valentin Gabeur, Yuan-Ting Hu, Ronghang Hu, Chaitanya Ryali,
  Tengyu Ma, Haitham Khedr, Roman R{\"a}dle, Chloe Rolland, Laura Gustafson,
  Eric Mintun, Junting Pan, Kalyan~Vasudev Alwala, Nicolas Carion, Chao-Yuan
  Wu, Ross Girshick, Piotr Doll{\'a}r, and Christoph Feichtenhofer.
\newblock {SAM 2}: {Segment Anything} in images and videos.
\newblock In \emph{International Conference on Learning Representations
  (ICLR)}, pages 28085--28128, 2025.

\bibitem[Ren et~al.(2023)Ren, He, Liu, Chen, Wang, Song, Li, Xue, Hu, Chen,
  Zheng, Xiang, and Bo]{ren2023make}
Jianqiang Ren, Chao He, Lin Liu, Jiahao Chen, Yutong Wang, Yafei Song, Jianfang
  Li, Tangli Xue, Siqi Hu, Tao Chen, Kunkun Zheng, Jianjing Xiang, and Liefeng
  Bo.
\newblock {Make-A-Character}: High quality text-to-{3D} character generation
  within minutes.
\newblock \emph{arXiv preprint arXiv:2312.15430}, 2023.

\bibitem[R{\"u}egg et~al.(2022)R{\"u}egg, Zuffi, Schindler, and
  Black]{rueegg2023barc}
Nadine R{\"u}egg, Silvia Zuffi, Konrad Schindler, and Michael~J. Black.
\newblock {BARC}: Learning to regress {3D} dog shape from images by exploiting
  breed information.
\newblock In \emph{IEEE/CVF Conference on Computer Vision and Pattern
  Recognition (CVPR)}, pages 3866--3874, 2022.
\newblock \doi{10.1109/CVPR52688.2022.00385}.

\bibitem[Seol et~al.(2011)Seol, Seo, Kim, Lewis, and Noh]{seol2011artist}
Yeongho Seol, Jaewoo Seo, Paul~Hyunjin Kim, J.~P. Lewis, and Junyong Noh.
\newblock Artist friendly facial animation retargeting.
\newblock \emph{ACM Transactions on Graphics}, 30\penalty0 (6):\penalty0 1--10,
  2011.
\newblock \doi{10.1145/2070781.2024196}.

\bibitem[Seol et~al.(2012)Seol, Lewis, Seo, Choi, Anjyo, and
  Noh]{seol2012spacetime}
Yeongho Seol, J.~P. Lewis, Jaewoo Seo, Byungkuk Choi, Ken Anjyo, and Junyong
  Noh.
\newblock Spacetime expression cloning for blendshapes.
\newblock \emph{ACM Transactions on Graphics}, 31\penalty0 (2):\penalty0
  14:1--14:12, 2012.
\newblock \doi{10.1145/2159516.2159519}.

\bibitem[Sharp et~al.(2022)Sharp, Attaiki, Crane, and
  Ovsjanikov]{sharp2022diffusionnet}
Nicholas Sharp, Souhaib Attaiki, Keenan Crane, and Maks Ovsjanikov.
\newblock {DiffusionNet}: Discretization agnostic learning on surfaces.
\newblock \emph{ACM Transactions on Graphics}, 41\penalty0 (3):\penalty0
  27:1--27:16, 2022.
\newblock \doi{10.1145/3507905}.

\bibitem[Sorkine and Alexa(2007)]{sorkine2007rigid}
Olga Sorkine and Marc Alexa.
\newblock As-rigid-as-possible surface modeling.
\newblock In \emph{Eurographics Symposium on Geometry Processing (SGP)}, pages
  109--116, 2007.
\newblock \doi{10.2312/SGP/SGP07/109-116}.

\bibitem[Sumner and Popovi{\'c}(2004)]{sumner2004deformation}
Robert~W. Sumner and Jovan Popovi{\'c}.
\newblock Deformation transfer for triangle meshes.
\newblock \emph{ACM Transactions on Graphics}, 23\penalty0 (3):\penalty0
  399--405, 2004.
\newblock \doi{10.1145/1015706.1015736}.

\bibitem[Tang et~al.(2024)Tang, Zeng, Fan, Wang, Dai, Chen, and
  Ma]{tang2024make}
Junshu Tang, Yanhong Zeng, Ke~Fan, Xuheng Wang, Bo~Dai, Kai Chen, and Lizhuang
  Ma.
\newblock {Make-It-Vivid}: Dressing your animatable biped cartoon characters
  from text.
\newblock In \emph{IEEE/CVF Conference on Computer Vision and Pattern
  Recognition (CVPR)}, pages 6243--6253, 2024.
\newblock \doi{10.1109/CVPR52733.2024.00597}.

\bibitem[Vlasic et~al.(2005)Vlasic, Brand, Pfister, and
  Popovi{\'c}]{vlasic2006face}
Daniel Vlasic, Matthew Brand, Hanspeter Pfister, and Jovan Popovi{\'c}.
\newblock Face transfer with multilinear models.
\newblock \emph{ACM Transactions on Graphics}, 24\penalty0 (3):\penalty0
  426--433, 2005.
\newblock \doi{10.1145/1073204.1073209}.

\bibitem[Wang et~al.(2015)Wang, Liu, Guo, Zhong, Le, and
  Deng]{wang2015spectral}
Chao Wang, Yang Liu, Xiaohu Guo, Zichun Zhong, Binh Le, and Zhigang Deng.
\newblock Spectral animation compression.
\newblock \emph{Journal of Computer Science and Technology}, 30\penalty0
  (3):\penalty0 540--552, 2015.
\newblock \doi{10.1007/s11390-015-1544-z}.

\bibitem[Wang et~al.(2026)Wang, Xiang, Sohn, Monnier, Yeh, Nguyen-Phuoc, Zhang,
  Fan, Toisoul, Jung, Dhar, Bunnell, Sarafianos, Zou, Shapovalov, Vedaldi, and
  Ranjan]{wang2026assetgen}
Dilin Wang, Xiaoyu Xiang, Kihyuk Sohn, Tom Monnier, Yu-Ying Yeh, Thu
  Nguyen-Phuoc, Jiawen Zhang, Yuchen Fan, Antoine Toisoul, Hyunyoung Jung,
  Prithviraj Dhar, Michael Bunnell, Nikolaos Sarafianos, Chuhang Zou, Roman
  Shapovalov, Andrea Vedaldi, and Rakesh Ranjan.
\newblock {AssetGen}: Deployable {3D} asset generation at interactive speed.
\newblock \emph{arXiv preprint arXiv:2605.26137}, 2026.

\bibitem[Wang et~al.(2023{\natexlab{a}})Wang, Li, Liu, De~Mello, Gallo, Wang,
  and Kautz]{wang2023zeroshot}
Jiashun Wang, Xueting Li, Sifei Liu, Shalini De~Mello, Orazio Gallo, Xiaolong
  Wang, and Jan Kautz.
\newblock Zero-shot pose transfer for unrigged stylized {3D} characters.
\newblock In \emph{IEEE/CVF Conference on Computer Vision and Pattern
  Recognition (CVPR)}, pages 8704--8714, 2023{\natexlab{a}}.
\newblock \doi{10.1109/CVPR52729.2023.00841}.

\bibitem[Wang et~al.(2023{\natexlab{b}})Wang, Qiu, Chen, Ding, and
  Pan]{wang2023fully}
Jingying Wang, Yilin Qiu, Keyu Chen, Yu~Ding, and Ye~Pan.
\newblock Fully automatic blendshape generation for stylized characters.
\newblock In \emph{IEEE Conference on Virtual Reality and 3D User Interfaces
  (VR)}, pages 347--355, 2023{\natexlab{b}}.
\newblock \doi{10.1109/VR55154.2023.00050}.

\bibitem[Wuu et~al.(2022)Wuu, Zheng, Ardisson, Bali, Belko, Brockmeyer, Evans,
  Godisart, Ha, Huang, Hypes, Koska, Krenn, Lombardi, Luo, McPhail,
  Millerschoen, Perdoch, Pitts, Richard, Saragih, Saragih, Shiratori, Simon,
  Stewart, Trimble, Weng, Whitewolf, Wu, Yu, and Sheikh]{wuu2022multiface}
Cheng-hsin Wuu, Ningyuan Zheng, Scott Ardisson, Rohan Bali, Danielle Belko,
  Eric Brockmeyer, Lucas Evans, Timothy Godisart, Hyowon Ha, Xuhua Huang,
  Alexander Hypes, Taylor Koska, Steven Krenn, Stephen Lombardi, Xiaomin Luo,
  Kevyn McPhail, Laura Millerschoen, Michal Perdoch, Mark Pitts, Alexander
  Richard, Jason Saragih, Junko Saragih, Takaaki Shiratori, Tomas Simon, Matt
  Stewart, Autumn Trimble, Xinshuo Weng, David Whitewolf, Chenglei Wu, Shoou-I
  Yu, and Yaser Sheikh.
\newblock {Multiface}: A dataset for neural face rendering.
\newblock \emph{arXiv preprint arXiv:2207.11243}, 2022.

\bibitem[Xie et~al.(2024)Xie, Xu, Song, Wang, Shi, and Luo]{xie2024xportrait}
You Xie, Hongyi Xu, Guoxian Song, Chao Wang, Yichun Shi, and Linjie Luo.
\newblock {X-Portrait}: Expressive portrait animation with hierarchical motion
  attention.
\newblock In \emph{ACM SIGGRAPH Conference Papers}, pages 1--11, 2024.
\newblock \doi{10.1145/3641519.3657459}.

\bibitem[Yang et~al.(2020)Yang, Zhu, Wang, Huang, Shen, Yang, and
  Cao]{yang2020facescape}
Haotian Yang, Hao Zhu, Yanru Wang, Mingkai Huang, Qiu Shen, Ruigang Yang, and
  Xun Cao.
\newblock {FaceScape}: A large-scale high quality {3D} face dataset and
  detailed riggable {3D} face prediction.
\newblock In \emph{IEEE/CVF Conference on Computer Vision and Pattern
  Recognition (CVPR)}, pages 598--607, 2020.
\newblock \doi{10.1109/CVPR42600.2020.00068}.

\bibitem[Yang et~al.(2024)Yang, Li, Wu, Jing, Li, Ji, Liang, and
  Fan]{yang2024megactor}
Shurong Yang, Huadong Li, Juhao Wu, Minhao Jing, Linze Li, Renhe Ji, Jiajun
  Liang, and Haoqiang Fan.
\newblock {MegActor}: Harness the power of raw video for vivid portrait
  animation.
\newblock \emph{arXiv preprint arXiv:2405.20851}, 2024.

\bibitem[Yu et~al.(2018)Yu, Wang, Peng, Gao, Yu, and Sang]{yu2018bisenet}
Changqian Yu, Jingbo Wang, Chao Peng, Changxin Gao, Gang Yu, and Nong Sang.
\newblock {BiSeNet}: Bilateral segmentation network for real-time semantic
  segmentation.
\newblock In \emph{European Conference on Computer Vision (ECCV)}, volume 11217
  of \emph{Lecture Notes in Computer Science}, pages 334--349. Springer, 2018.
\newblock \doi{10.1007/978-3-030-01261-8_20}.

\bibitem[Zhang et~al.(2025)Zhang, Wang, Zhang, Xu, Song, Xie, Luo, Tian, Feng,
  and Guo]{zhang2025magictalk}
Chenxu Zhang, Chao Wang, Jianfeng Zhang, Hongyi Xu, Guoxian Song, You Xie,
  Linjie Luo, Yapeng Tian, Jiashi Feng, and Xiaohu Guo.
\newblock {MagicTalk}: Implicit and explicit correlation learning for
  diffusion-based emotional talking face generation.
\newblock \emph{Computational Visual Media}, 11\penalty0 (4):\penalty0
  763--779, 2025.
\newblock \doi{10.26599/CVM.2025.9450491}.

\bibitem[Zhang et~al.(2023)Zhang, Qiu, Lin, Zhang, Shi, Yang, Shi, Yang, Xu,
  and Yu]{zhang2023dreamface}
Longwen Zhang, Qiwei Qiu, Hongyang Lin, Qixuan Zhang, Cheng Shi, Wei Yang,
  Ye~Shi, Sibei Yang, Lan Xu, and Jingyi Yu.
\newblock {DreamFace}: Progressive generation of animatable {3D} faces under
  text guidance.
\newblock \emph{ACM Transactions on Graphics}, 42\penalty0 (4):\penalty0
  138:1--138:16, 2023.
\newblock \doi{10.1145/3592094}.

\bibitem[Zhu and Joslin(2024)]{zhu2024fabrig}
ChangAn Zhu and Chris Joslin.
\newblock {Fabrig}: A cloth-simulated transferable {3D} face parameterization.
\newblock In \emph{ACM SIGGRAPH Asia Conference Papers}, pages 111:1--111:10,
  2024.
\newblock \doi{10.1145/3680528.3687695}.

\bibitem[Zhu et~al.(2023)Zhu, Yang, Guo, Zhang, Wang, Huang, Wu, Shen, Yang,
  and Cao]{zhu2023facescape}
Hao Zhu, Haotian Yang, Longwei Guo, Yidi Zhang, Yanru Wang, Mingkai Huang,
  Menghua Wu, Qiu Shen, Ruigang Yang, and Xun Cao.
\newblock {FaceScape}: {3D} facial dataset and benchmark for single-view {3D}
  face reconstruction.
\newblock \emph{IEEE Transactions on Pattern Analysis and Machine
  Intelligence}, 45\penalty0 (12):\penalty0 14528--14545, 2023.
\newblock \doi{10.1109/TPAMI.2023.3307338}.

\bibitem[Zuffi et~al.(2017)Zuffi, Kanazawa, Jacobs, and
  Black]{zuffi2017menagerie}
Silvia Zuffi, Angjoo Kanazawa, David~W. Jacobs, and Michael~J. Black.
\newblock {3D Menagerie}: Modeling the {3D} shape and pose of animals.
\newblock In \emph{IEEE Conference on Computer Vision and Pattern Recognition
  (CVPR)}, pages 5524--5532, 2017.
\newblock \doi{10.1109/CVPR.2017.586}.

\end{thebibliography}

\clearpage
\appendix
\setcounter{figure}{0}
\setcounter{table}{0}
\setcounter{equation}{0}
\numberwithin{figure}{section}
\numberwithin{table}{section}
\numberwithin{equation}{section}
\renewcommand{\theHsection}{appendix.\Alph{section}}
\renewcommand{\theHfigure}{\theHsection.\arabic{figure}}
\renewcommand{\theHtable}{\theHsection.\arabic{table}}
\renewcommand{\theHequation}{\theHsection.\arabic{equation}}
\newcommand{\supprev}[1]{#1}

\section*{\huge Appendix}
\addcontentsline{toc}{section}{Appendix}

\section{Supplementary Riggability Assessment Details}
\label{supp:sec:riggability}

This section expands the design of the hybrid riggability checker introduced in the main paper. We give the per-question signal assignment, the geometric criteria used to validate segmentation masks, the landmark-based mouth-quality gate, the mouth-signal fusion logic, and a brief discussion of the remaining failure modes that bound the accuracy of the gate.

\subsection{VLM+CV Signal Architecture}

The checker fuses one semantic signal and two CV-derived signals computed from a single rendered view of the input asset:

\begin{itemize}
    \item \textbf{VLM semantic signal.} \supprev{Gemini 2.5 Flash Image is the primary vision-language model, with a Llama~4 multimodal model as fallback when the primary response is unavailable or invalid.} It answers a fixed set of high-level questions about the rendered character (face type, number of faces, eye orbits, accessories, mouth state, and teeth visibility). These questions capture failure modes that are inherently semantic and hard to phrase as geometric thresholds.
    \item \textbf{Segmentation-mask signal.} The face/eye/mouth masks produced by the segmentation ensemble are evaluated against geometric plausibility criteria: presence of each region, inclusion inside the face boundary, plausible eye-to-head ratio, and acceptable lip-seal curvature.
    \item \textbf{Landmark-quality signal.} For human-like faces, the regularity of mouth landmarks is summarized by a lip-evenness statistic and used as a quality gate before mouth evidence enters the fusion step.
\end{itemize}

The three signals are not treated as redundant: each question in the underlying checker template is assigned to the signal that is empirically most reliable for it, as summarized in Table~\ref{supp:tab:signal-assignment}.

\begin{table}[htbp]
\centering
\caption{Per-question signal assignment in the riggability checker. ``VLM'' uses the semantic signal alone; ``Seg mask'' replaces a VLM yes/no question with a direct geometric measurement; ``Majority'' fuses VLM, segmentation, and landmark evidence with a three-way vote.}
\label{supp:tab:signal-assignment}
\small
\begin{tabular}{@{}lll@{}}
\toprule
Q\# & Aspect checked & Signal source \\
\midrule
Q1  & Face type (human / animal / other) & VLM \\
Q2  & Number of visible faces & VLM \\
Q3  & Number of visible eye orbits & VLM \\
Q4  & Mouth present and localizable & Majority \\
Q5  & Eye-region occlusion or obstruction & Seg mask \\
Q6  & Eyewear covering eyes & VLM \\
Q7  & Mouth state & VLM \\
Q8  & Teeth visibility & VLM \\
Q9  & Eye-to-head size ratio & Seg mask \\
Q10 & Lip-seal curvature & Seg mask \\
\bottomrule
\end{tabular}
\end{table}

\subsection{Geometric Criteria from Segmentation Masks}

The segmentation-based checks (Q5, Q9, Q10, plus mask inclusion tests) are summarized in Table~\ref{supp:tab:geometric-criteria}. Their binary decision rules are calibrated once on a development split and held fixed across all evaluations; they are deliberately conservative so that the checker mostly rejects inputs that would otherwise crash or distort the downstream registration stage.

\begin{table}[htbp]
\centering
\caption{Geometric criteria evaluated on the face, eye, and mouth segmentation masks. Each criterion produces a binary pass/fail signal that participates in the hybrid decision.}
\label{supp:tab:geometric-criteria}
\small
\begin{tabular}{@{}p{0.36\columnwidth}p{0.55\columnwidth}@{}}
\toprule
Criterion & Pass condition \\
\midrule
Mask presence & face/eye/mouth masks non-empty \\
Eye inclusion & eye mask inside face boundary \\
Mouth inclusion & mouth mask inside face boundary \\
Eye-region occlusion (Q5) & occluder overlap with eye region below threshold \\
Eye-to-head ratio (Q9) & eye/head area ratio in range \\
Mouth curvature (Q10) & lip-seal curvature in range \\
\bottomrule
\end{tabular}
\end{table}

\subsection{Landmark-Assisted Mouth Validation}

For human-like inputs, we use mouth landmarks as a quality gate before they are fused with the VLM and segmentation signals. Let $\kappa_{\mathrm{lip}}$ denote the coefficient of variation (standard deviation divided by mean) of the spacing between consecutive lip-contour landmarks. We mark the mouth landmarks as unreliable when $\kappa_{\mathrm{lip}}$ exceeds a calibrated threshold; the underlying intuition is that on assets with a poorly modeled or absent mouth, the landmark detector still emits points, but those points become irregularly spaced. Failing the gate sets the landmark mouth evidence to ``unavailable'' for the subsequent fusion step.

For non-human assets, where human-trained mouth-landmark detectors are not reliable, the landmark signal is ignored and the mouth decision is determined by the VLM and the segmentation mask alone.

\subsection{Mouth-Signal Fusion Logic}

The final eligibility decision is the conjunction of all per-question results in Table~\ref{supp:tab:signal-assignment}. Q1--Q3 and Q6--Q8 are decided directly by the VLM; Q5, Q9, Q10 are decided by the segmentation-mask checks in Table~\ref{supp:tab:geometric-criteria}; and Q4 (mouth presence) uses explicit fusion because mouth localization is one of the most common downstream failure sources.

Concretely, for a human-like asset, the mouth decision is:

\begin{enumerate}
    \item If $\kappa_{\mathrm{lip}}$ exceeds the landmark-quality threshold, treat the landmark evidence as unreliable and rely on the VLM and segmentation signals.
    \item Otherwise, take a majority vote over (i) the VLM mouth prediction, (ii) the segmentation-based mouth-detected flag, and (iii) the landmark-based mouth-detected flag.
\end{enumerate}

For non-human assets, the same step degenerates to a conjunction of the VLM and segmentation-mask signals. The majority-vote design recovers the cases that a strict ``VLM and segmentation'' rule would miss when exactly one of the three signals is wrong but the other two agree.

\subsection{Discussion of Remaining Failure Modes}
\label{supp:sec:failure_modes}

Empirically, the hybrid VLM+CV checker on the mixed riggability validation set exhibits two dominant failure categories. The first is \emph{annotation borderline cases}, where the ground-truth label itself is ambiguous (e.g., partially occluded faces or stylized characters that are arguably riggable). The second is \emph{VLM-only semantic errors}: misclassification of eyewear, mouth state, or teeth visibility---attributes that segmentation masks and landmarks only partially corroborate. These are the natural targets for future improvement of the checker, since the geometric and landmark signals have already absorbed most of the failures they are capable of catching.

\section{Supplementary Stage-2 Rig Construction Details}
\label{supp:sec:stage2}

This section expands the four implementation components underlying the three Stage-2 steps in the main paper: face-region fusion; inner-mouth construction; texture/UV transfer; and blendshape transfer with post-processing. The shared difficulty across these components is that the input asset is surface-only: there is no pre-modeled oral cavity, no animation-grade facial topology, and no separation between expressive and structural regions.

\subsection{Face-Region Fusion}

The fitted face mesh produced by Stage 1 defines the region of the input asset that will be replaced. A high-quality animated topology---designed with radial edge loops around the eyes and mouth---is first aligned to the fitted template through their shared UV correspondence, then reprojected onto the input surface so that it inherits the asset's own facial geometry. The original face region of the input is removed via a Boolean operation against a convex hull of the fitted region, the new animated topology is inserted in its place, and the boundary between the new face and the rest of the asset is stitched by closest-point welding followed by small-edge collapse to clean up degenerate polygons. This operation is intentionally local: it preserves the asset's identity, hairstyle, and non-face geometry while introducing the radial edge flow that downstream expression deformation relies on.

\subsection{Inner-Mouth Construction}

Because the input contains no oral cavity, the inner-mouth geometry (teeth, gums, and tongue) cannot be recovered from the asset itself. We instead maintain a small library of inner-mouth template archetypes, summarized in Table~\ref{supp:tab:archetypes}; the appropriate archetype is selected by the VLM during Stage 1 and fitted to the character as described below.

\begin{table}[htbp]
\centering
\caption{Inner-mouth template archetypes and their typical targets. The VLM in Stage 1 selects one archetype per asset, which determines the teeth/gum/tongue template that is warped into the fitted mouth cavity.}
\label{supp:tab:archetypes}
\small
\begin{tabular}{@{}p{0.20\columnwidth}p{0.69\columnwidth}@{}}
\toprule
Archetype & Typical target characters \\
\midrule
Human & realistic humans and humanoid characters \\
Canine & long-muzzled animals (e.g., dogs, wolves, foxes) \\
Monster & other non-human characters with non-standard teeth (e.g., cats, bears, gorillas, and less-common stylized species) \\
\bottomrule
\end{tabular}
\end{table}

\supprev{Each face template defines an inner-mouth region, and every teeth archetype is spatially registered to that region at rest. After the character-specific face template is fitted to the target surface, its inner-mouth region is deformed as-rigid-as-possible (ARAP) to the new shape. We then estimate, by SVD, the rigid transformation between the canonical and deformed inner-mouth regions and apply it to the selected teeth, gum, and tongue archetype. A convex hull of the aligned teeth provides the collision proxy. At rest, a point-to-SDF penalty moves intersecting outer-face points outside this hull; the same penalty is applied again during per-expression deformation transfer. This treatment is particularly important for long-muzzled animals, where initial collisions may otherwise occur.}

\subsection{Texture and UV Transfer}

After the new face and inner-mouth surfaces are inserted, the global UV layout is rebuilt. The face region is allocated a larger fraction of the texture domain than non-face regions so that high-frequency facial detail is preserved at any rendering resolution, and the inner-mouth surfaces are placed on a separate texture channel to keep their material parameters independent. The asset's original appearance is then transferred onto the rebuilt UV layout by sampling the original texture through the geometric correspondence between the original and merged surfaces; this preserves the character's identity and material details while letting the newly synthesized geometry inherit a coherent appearance.

\subsection{Blendshape Transfer and Post-Processing}

FACS expression shapes are produced by transferring local deformation gradients from a canonical rigged template to the merged character mesh using sparse deformation transfer. Adapting the deformation gradients to the target character's own proportions is what keeps the transferred expressions character-specific rather than collapsing the rig toward an ``average human face,'' and is what allows the same set of FACS shapes to be applied to humans, humanoids, long-muzzled animals, and short-muzzled animals without per-character authoring.

The transferred deformations are then blended outward from the facial template region into the adjacent head surface so that expression motion fades smoothly into the rest of the asset. Sparse deformation transfer alone cannot enforce certain physical and geometric constraints---eyelids must close without self-intersection, eye-gaze shapes require a known rotation center, jaw motion must move the lower teeth, and extreme mouth poses can re-introduce teeth--face intersections. A small set of expression-specific post-processing rules, summarized in Table~\ref{supp:tab:post-processing}, handles these cases.

\begin{table}[htbp]
\centering
\caption{Expression-specific post-processing rules applied after sparse deformation transfer. Each rule addresses a constraint that pure deformation transfer cannot enforce on its own.}
\label{supp:tab:post-processing}
\small
\begin{tabular}{@{}p{0.38\columnwidth}p{0.53\columnwidth}@{}}
\toprule
Shape category & Refinement applied \\
\midrule
Closed-eye shapes & upper-eyelid vertices snapped to lower lid and locally relaxed for contact-aware closure \\
Eye-gaze shapes & eye-region vertices rotate around virtual eyeball centers from Procrustes fit \\
Jaw-related shapes & lower-teeth subset moves coherently with the jaw, tracking open/close motion \\
    Collision-aware refinement & \supprev{point-to-SDF teeth-hull penalty reapplied per expression shape} \\
\bottomrule
\end{tabular}
\end{table}

\subsection{Design Tradeoffs and Outlook}

The four implementation components of Stage 2 are coupled but decomposable: face-region fusion fixes the surface topology, inner-mouth construction synthesizes the missing oral cavity, UV/texture transfer preserves the character's original appearance, and blendshape transfer turns the static rig into an animatable one. A common thread is that each component substitutes a procedural geometric operation for what is traditionally a manual authoring step, which is also what makes the full pipeline fully automatic.

The two design choices that most directly enable cross-topology generalization are the use of multiple inner-mouth archetypes (Table~\ref{supp:tab:archetypes}) and the SDF-based collision refinement at both the rest pose and the per-expression level. Without the archetype split, a single human-shaped teeth template either intersects long muzzles or sits incorrectly in animals with non-human teeth structure (e.g., cats, bears, gorillas); without the SDF refinement, expression shapes such as a wide jaw open or a strong smile can re-introduce intersections that the rest-pose placement avoids. Together with the topology-specific face templates of Stage 1, these choices let the same Stage 2 procedure produce rigs for humans, humanoids, long-muzzled animals, and short-muzzled animals \supprev{without per-asset manual authoring}.

Together, these components turn a static, surface-only 3D character mesh---without any pre-modeled oral cavity, without any animation-grade facial topology, and without any per-character authored expressions---into a full inner-mouth-aware FACS facial rig.

\section{Additional Qualitative Results}

This section provides additional qualitative result grids beyond \Cref{fig:qualitative_results} in the main paper. Each grid follows the same controlled layout as the main qualitative figure: the leftmost column shows the original static surface-only asset, and the remaining six columns show generated expressions with inner-mouth geometry from the OmniFaceRig output. These additional results demonstrate robustness across more identities, styles, and topology families without overloading the main paper.

\clearpage

  \begin{figure}[p]
    \centering
    \includegraphics[width=0.97\textwidth]{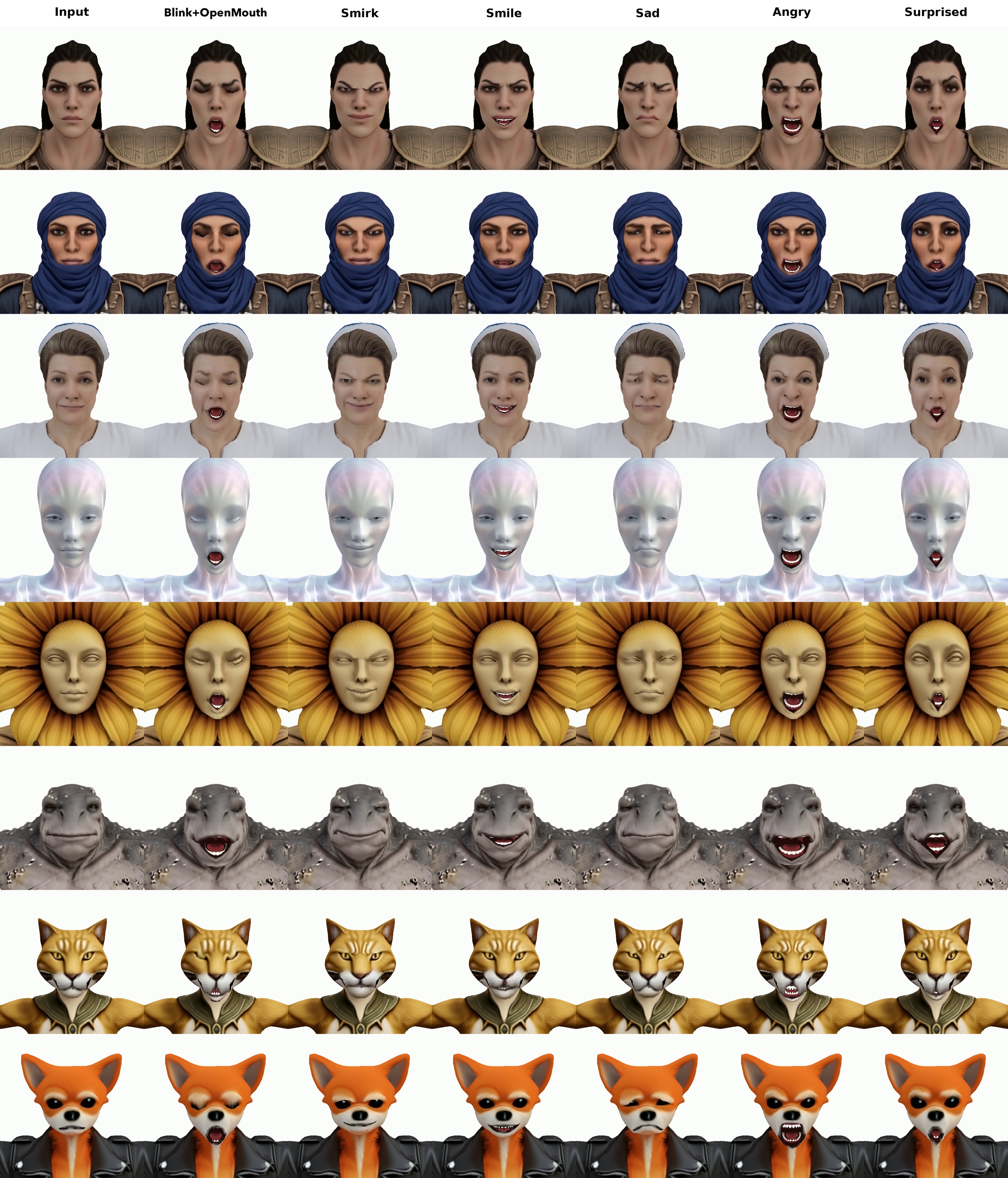}
    \caption{Additional qualitative OmniFaceRig results (1). Each row shows one surface-only input asset (left), followed by six generated expressions with inner-mouth geometry.}
    \label{supp:fig:qualitative-000}
  \end{figure}
  \clearpage
  \begin{figure}[p]
    \centering
    \includegraphics[width=0.97\textwidth]{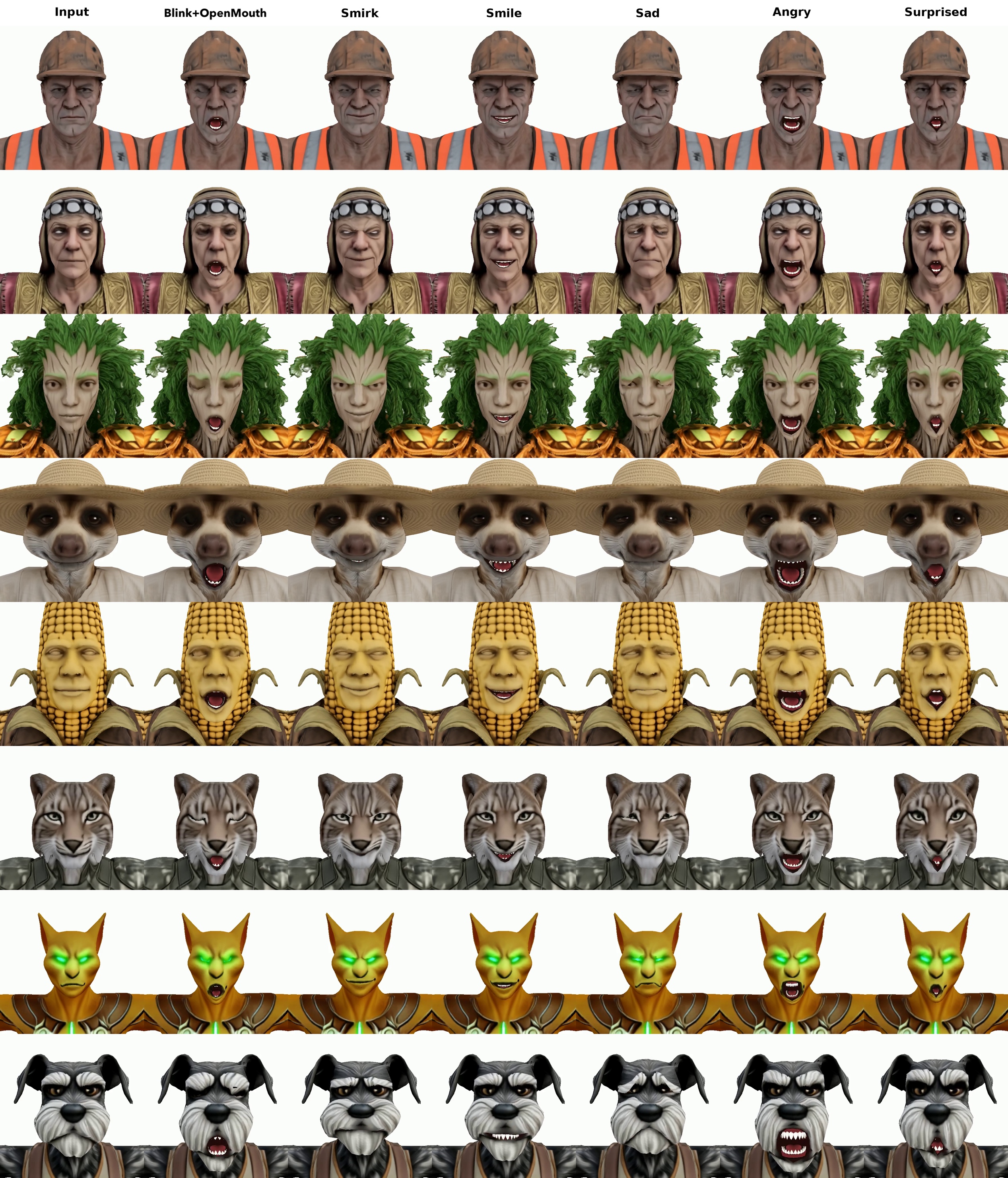}
    \caption{Additional qualitative OmniFaceRig results (2). Each row shows one surface-only input asset (left), followed by six generated expressions with inner-mouth geometry.}
    \label{supp:fig:qualitative-001}
  \end{figure}
  \clearpage
  \begin{figure}[p]
    \centering
    \includegraphics[width=0.97\textwidth]{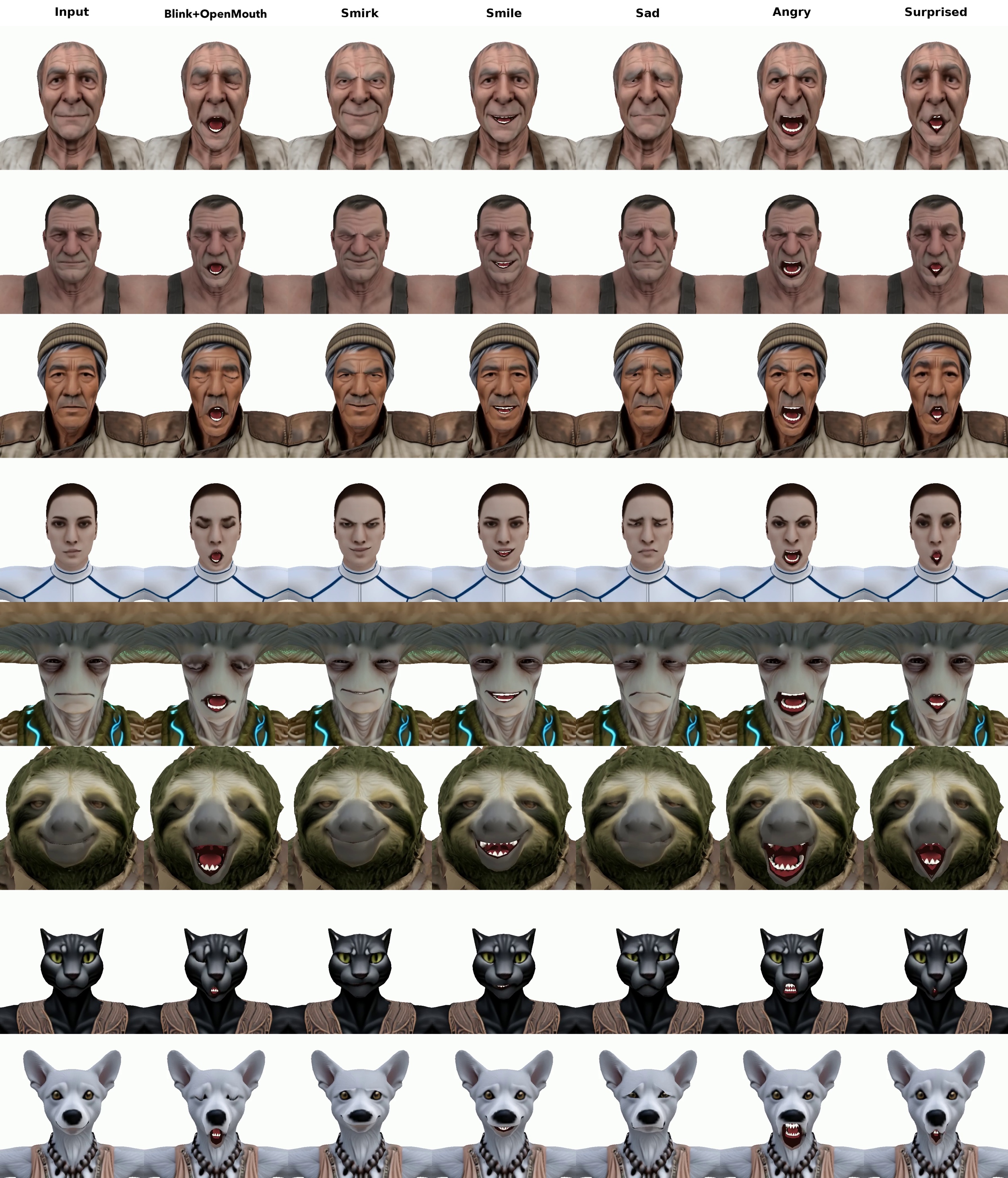}
    \caption{Additional qualitative OmniFaceRig results (3). Each row shows one surface-only input asset (left), followed by six generated expressions with inner-mouth geometry.}
    \label{supp:fig:qualitative-002}
  \end{figure}
  \clearpage
  \begin{figure}[p]
    \centering
    \includegraphics[width=0.97\textwidth]{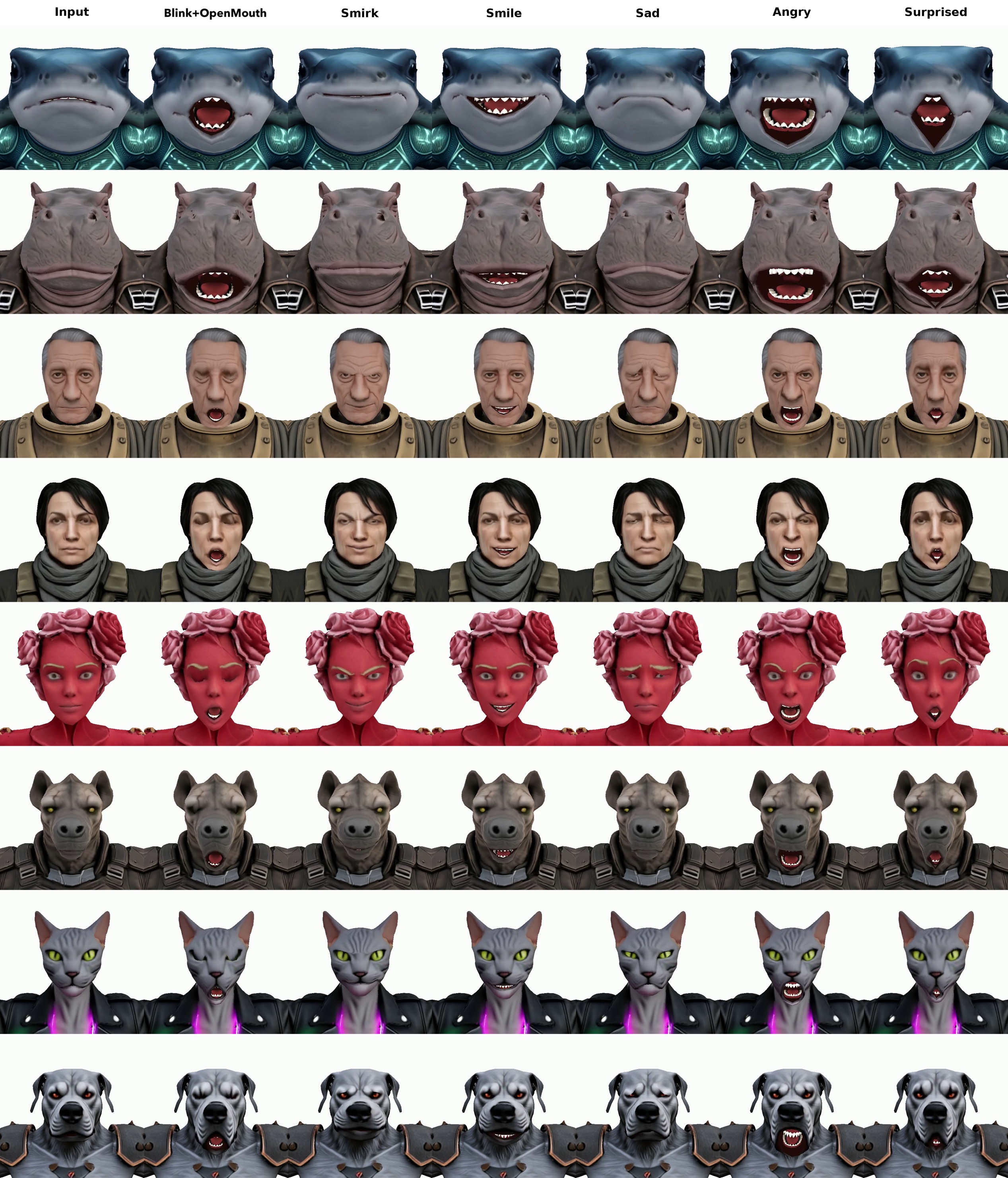}
    \caption{Additional qualitative OmniFaceRig results (4). Each row shows one surface-only input asset (left), followed by six generated expressions with inner-mouth geometry.}
    \label{supp:fig:qualitative-003}
  \end{figure}
  \clearpage
  \begin{figure}[p]
    \centering
    \includegraphics[width=0.97\textwidth]{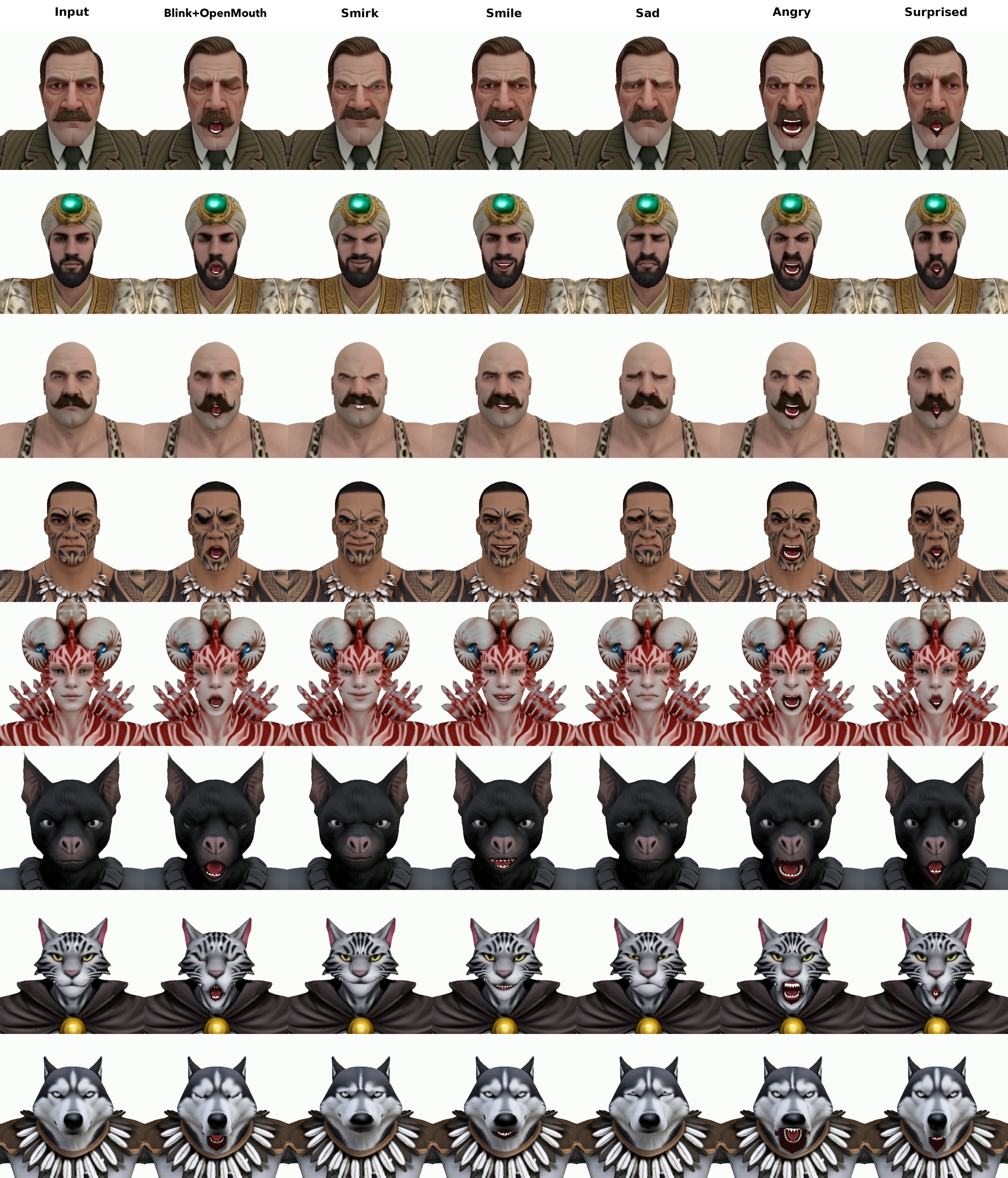}
    \caption{Additional qualitative OmniFaceRig results (5). Each row shows one surface-only input asset (left), followed by six generated expressions with inner-mouth geometry.}
    \label{supp:fig:qualitative-004}
  \end{figure}
  \clearpage
  \begin{figure}[p]
    \centering
    \includegraphics[width=0.97\textwidth]{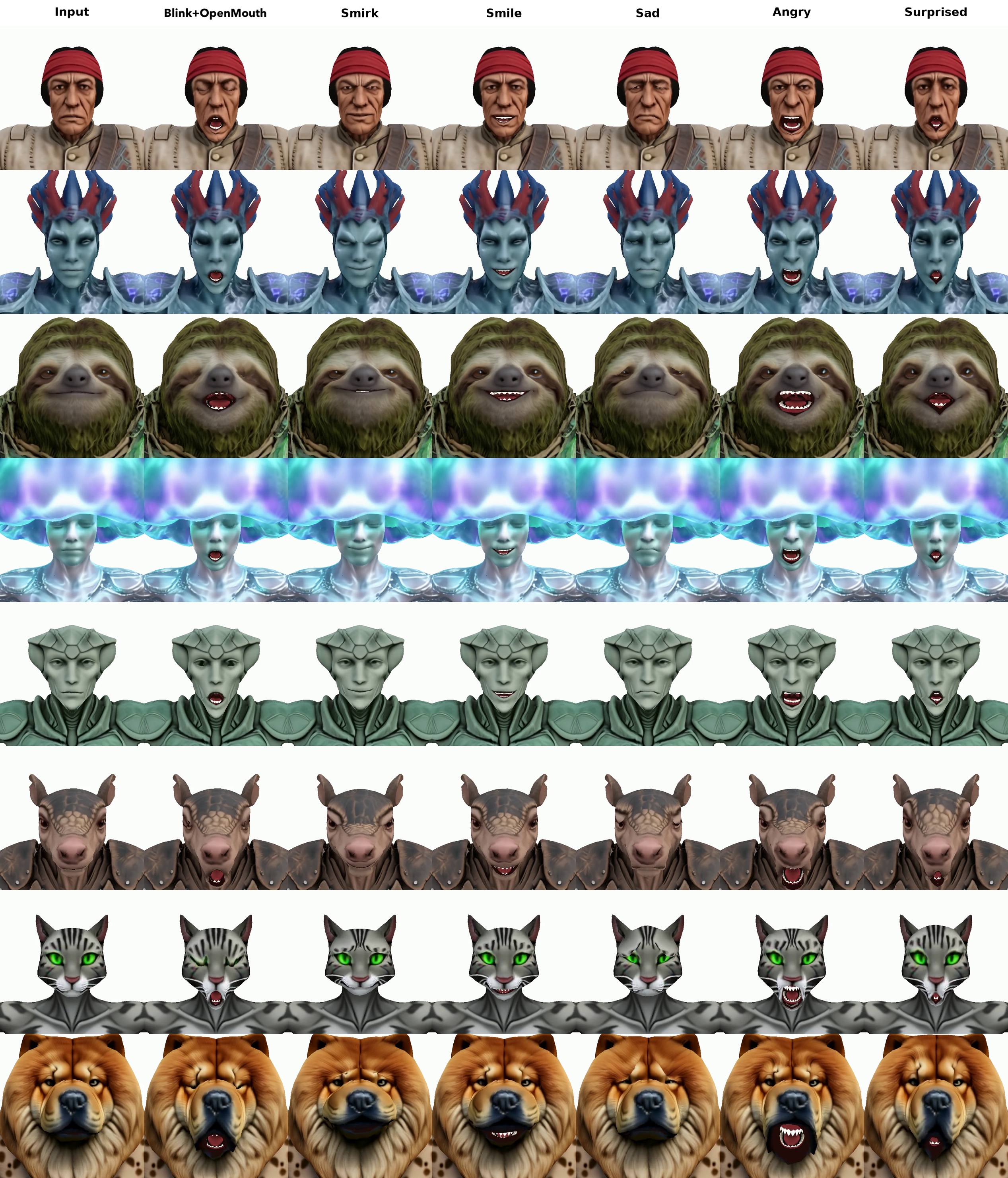}
    \caption{Additional qualitative OmniFaceRig results (6). Each row shows one surface-only input asset (left), followed by six generated expressions with inner-mouth geometry.}
    \label{supp:fig:qualitative-005}
  \end{figure}
  \clearpage
  \begin{figure}[p]
    \centering
    \includegraphics[width=0.97\textwidth]{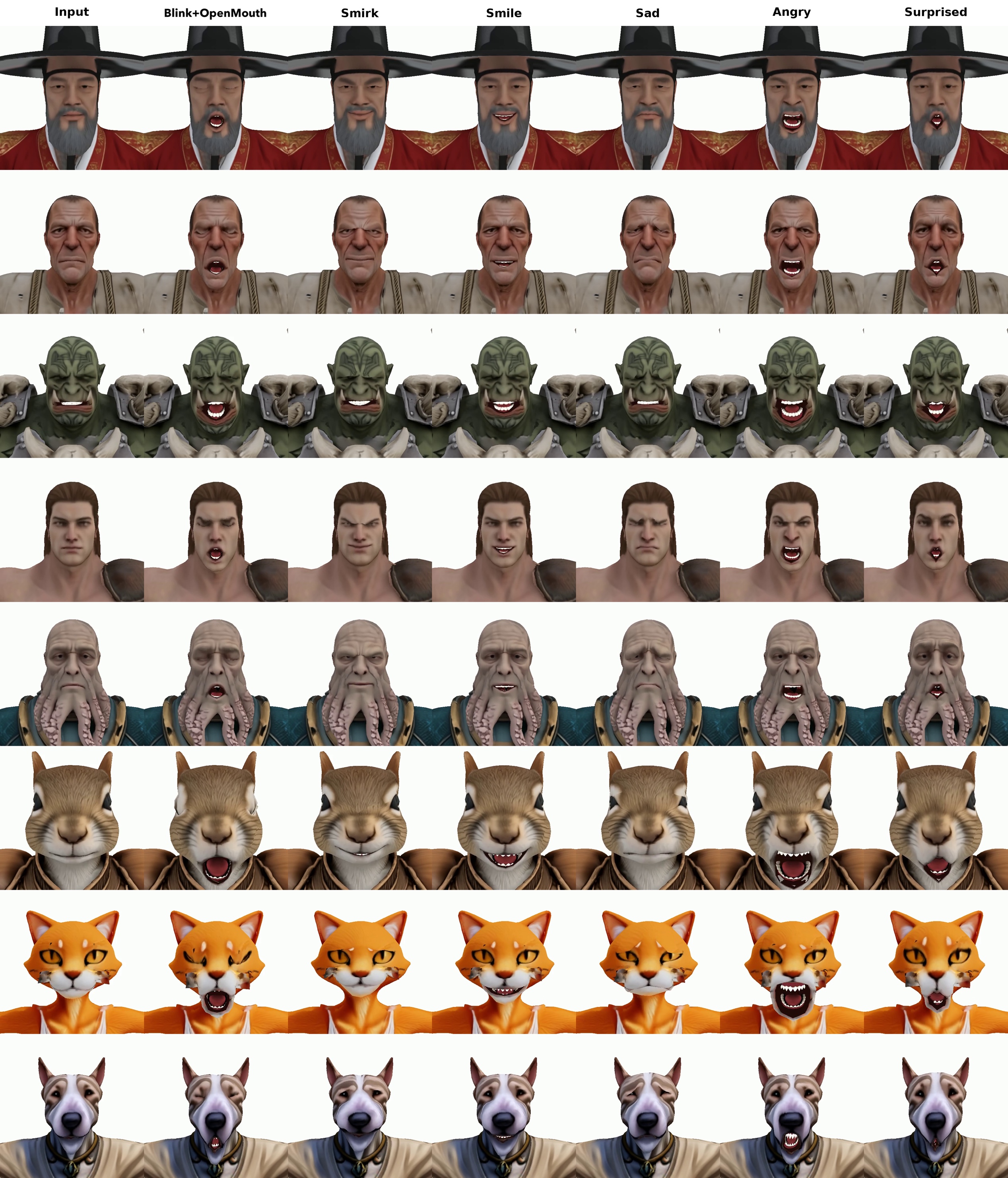}
    \caption{Additional qualitative OmniFaceRig results (7). Each row shows one surface-only input asset (left), followed by six generated expressions with inner-mouth geometry.}
    \label{supp:fig:qualitative-006}
  \end{figure}
  \clearpage

\end{document}